\documentclass[12pt,epsf]{article}  
\usepackage{amsmath}  

\usepackage{amsmath,amsfonts,latexsym,amssymb}
\usepackage{mathrsfs}

\def\ben{\begin{equation}}
\def\een{\end{equation}}
\def\bena{\begin{eqnarray}}
\def\eena{\end{eqnarray}}
\def\half{{\frac{1}{2}}}

\def\f(#1/#2){\frac{#1}{#2}} 
\def\Frac(#1/#2){\left(\frac{#1}{#2}\right)} 
\def\chris(#1-#2-#3){{\mit \Gamma}^{#1}{}_{{#2}{#3}} }
\def\tilchris(#1-#2-#3){\tilde{{\mit \Gamma}}^{#1}{}_{{#2}{#3}}}
\def\hatchris(#1-#2-#3){\hat{{\mit \Gamma}}^{#1}{}_{{#2}{#3}}}

\def\tilR{\tilde{R}}

\newcommand{\non}{\nonumber}

\newcommand{\dd}{{\rm d}}
\newcommand{\I}{{\mathscr I}}

\newcommand{\tC}{{\tilde C}}
\newcommand{\tg}{{\tilde g}}
\newcommand{\tf}{{\tilde f}}
\newcommand{\tell}{{\tilde \ell}}
\newcommand{\tn}{{\tilde n}}
\newcommand{\tnabla}{{\tilde \nabla}}

\newcommand{\tS}{{\tilde S}}

\newcommand{\e}{{\rm e}}

\begin{document}
%-----------------------------------------------------------------------------

\title{
Higher Dimensional Bondi Energy with a Globally Specified 
Background Structure{}\thanks{KEK-Cosmo-3 \quad KEK/TH/1213} 
}

\author{Akihiro Ishibashi{}\thanks{\tt akihiro.ishibashi@kek.jp}\: 
\\ \\
%%%
{\it Cosmophysics Group, Institute of Particle and Nuclear Studies} \\
{\it KEK, Tsukuba, Ibaraki, 305-0801, Japan} \\
  and \\  
%%%
{\it Enrico Fermi Institute, } \\
{\it The University of Chicago, Chicago, IL 60637, USA} \\ 
        }

\maketitle

%---------------------------------------------------------------------------
\begin{abstract}
A higher (even spacetime) dimensional generalization of 
the Bondi energy has recently been proposed \cite{HI05} 
within the framework of conformal infinity and Hamiltonian formalizm.  
The gauge condition employed in \cite{HI05} to derive the Bondi energy 
expression is, however, peculiar in the sense 
that cross-sections of null infinity specified by that gauge 
are anisotropic and in fact non-compact. 
For this reason, 
% it is not convenient to take that gauge when one 
that gauge is difficult to use for explicit computation of the Bondi energy 
in general, asymptotically flat radiative spacetimes. 
Also it is not clear, under that gauge condition, whether apparent 
difference between the expressions of higher dimensional Bondi energy 
and the $4$-dimensional one is due to the choice of gauges or 
qualitatively different nature of higher dimensional gravity 
from $4$-dimensional gravity. 
In this paper, we consider instead, Gaussian null conformal gauge as one of 
more natural gauge conditions that admit a global specification of 
background structure with compact, spherical cross-sections of null infinity. 
Accordingly, we modify the previous definition of 
higher dimensional news tensor so that it becomes well-defined 
in the Gaussian null conformal gauge and derive, for vacuum solutions, 
an expression for the Bondi energy-momentum in the new gauge choice, 
which takes a universal form in arbitrary (even spacetime) dimensions 
greater than or equal to four.  

\medskip 
\noindent 
PACS number: 04.20. -q, 04.20. Ha, 04.20. Fy, 04.50.-h 
\end{abstract} 

\newpage 

\section{Introduction}
%---------------------------------------------------------------------------
There has recently been a considerable interest in theories formulated in 
higher dimensional spacetimes, and therefore it is of great importance 
to define a precise notion of the total energy of an isolated gravitating 
system in higher dimensions. 
%In general relativity an isolated gravitating system is described 
%in terms of the notion of asymptotic flatness, which makes it possible 
%to define the total energy of the system in a satisfactory way. 
% A very useful treatment of isolated gravitating systems 
% in general relativity is the notion of asymptotic flatness, which makes 
In $4$-dimensional general relativity, there are two distinguished  
notions of asymptotic flatness and an associated total energy: 
The ADM energy~\cite{ADM} defined at spatial infinity and 
the Bondi energy~\cite{Trautman58,BBM} defined at a cross section of 
null infinity. 
The former is a constant in time, while the latter is 
in general, time-dependent and can be used to measure energy loss of 
an isolated source by emission of radiation.  
In this paper, we are concerned with the latter energy in higher dimensions. 

\medskip 
S.~Hollands and the present author have previously proposed a higher 
(even spacetime) dimensional generalization of the Bondi energy~\cite{HI05}. 
We first studied conditions for asymptotic flatness 
at null infinity $\I$ in higher dimensions within the framework of conformal 
infinity, analysing the stability of conformal null infinity 
against linear gravitational perturbations. 
Then we derived an expression for the generator conjugate to 
asymptotic symmetries within the Hamiltonian framework of Wald and 
Zoupas~\cite{Wald-Zoupas00} and proposed to take that generator conjugate 
to asymptotic time translation as the definition of a higher dimensional 
Bondi energy. In the derivation, we employed a particular gauge 
which demands that a conformal background geometry be locally, exactly 
Minkowskian in some neighborhood of conformal null infinity~$\I$. 
% (see eq.~(59) of \cite{HI05}). 
(We shall review below the gauge choice in \cite{HI05}, which we refer to 
as the {\sl Minkowskian conformal gauge} in this paper.) 
This gauge choice simplifies relevant computations to certain extent 
and thus is convenient for writing down the symplectic potential---whose 
integral over a segment of $\I$ corresponds to radiation flux passing 
through the segment---as well as deriving 
the Bondi energy expression in terms of curvature tensor with respect to 
unphysical conformal metric. 
However the Minkowskian conformal gauge is not globally well-defined 
in the sense that under that gauge, cross-sections of conformal null infinity 
$\I$, where we evaluate the Bondi-energy, become non-compact; 
they are sphere with a single point removed and do not naturally 
reflect the topology of $\I$. 
For the purpose of computing the Bondi-energy in general, asymptotically flat 
radiative spacetimes and obtaining physical consequences,  
it would be much preferable that the Bondi-energy is expressed 
in gauges that can be taken {\sl globally} in a neighborhood 
of $\I$ so that the Bondi-energy is evaluated on a {\sl compact}  
cross-section of $\I$. 
Also, with the Minkowskian conformal gauge, the definition of higher 
dimensional news tensor and the Bondi energy expression \cite{HI05} are 
slightly different from their $4$-dimensional counter parts 
obtained previously~\cite{Geroch77,Wald-Zoupas00}. 
It is not clear whether the differences are due to the choice of a peculiar 
gauge or qualitatively different nature of higher dimensional gravity from 
$4$ dimensional one. 

\medskip 
The purpose of this paper is to obtain an expression for the Bondi-energy 
in even-spacetime dimensions under a natural gauge choice that allows us to 
take compact, spherical cross-sections of $\I$. 
% 
% Specifically, we shall consider Gaussian null coordinates 
% in a neighborhood of $\I$ in an unphysical, 
% conformal spacetime $(\tilde M,\tilde g_{ab})$.   
% which we refer to as {\sl Gaussian null conformal gauge}. 
%
In even spacetime dimensions greater than or equal to four, 
a conformal null infinity $\I$ can exist as a smooth boundary null 
hypersurface of an unphysical, conformal spacetime $(\tilde M,\tilde g_{ab})$, 
being regular even in the presence of radiation~\cite{HI05}, in contrast to 
the case of odd spacetime dimensions~\cite{Hollands-Wald04}. 
(For discussions on smoothness of null infinity in $4$-dimensions, 
see e.g., \cite{CW89,CD02,Friedrich03} and references therein.) 
Following the standard arguments, one can then construct
%, without loss of generality, 
Gaussian null coordinates---which we refer to as 
{\sl Gaussian null conformal gauge}---in a neighborhood of $\I$ 
on the unphysical, conformal spacetime. 
Therefore, for even dimensions, in addition to the drop off conditions 
on the conformal metric, $\tg_{ab}$, derived in \cite{HI05}, 
one can naturally assume, as a part of the definition of asymptotic 
flatness, that the Gaussian null conformal gauge can be taken 
in some neighborhood of conformal null infinity. 
We discuss that the Gaussian null conformal gauge covers $\I$ globally 
and naturally specifies compact, spherical cross sections of $\I$. 

\medskip 
In the next section, we briefly summarize the main results of 
the paper~\cite{HI05}. We first recapitulate our definition of asymptotic 
flatness at null infinity in higher (even) dimensions and our strategy 
for deriving a Bondi-type energy in higher dimensions within 
the Hamiltonian framework. Then, we recall our higher dimensional 
Bondi-energy expression defined under the Minkowskian conformal gauge. 
In Section~\ref{sect:GNC}, we introduce Gaussian null coordinates 
in a neighborhood of null infinity in an unphysical, conformal spacetime 
and express our asymptotic flatness conditions in terms of that coordinate 
system with respect to $\I$. 
%
% \footnote{ %%% 
This Gaussian null coordinate system is useful to see differences 
between asymptotic behavior of higher dimensional gravity and that of 
$4$-dimensional gravity. For example, in $4$-dimensions, it is well-known 
that asymptotic symmetry group at null infinity is not the Poincare group 
but rather an infinite dimensional group, called Bondi-Metzner-Sachs (BMS) 
group, which includes angle-dependent translations or 
{\sl supertranslations}~\cite{BBM,Sachs62b,Geroch77}. 
On the other hand, it has been shown in~\cite{HI05} that supertranslation 
does not exist in higher dimensions. This can be manifest and easy to 
compare with the $4$-dimensional case when one examines asymptotic 
translational symmetries in terms of Gaussian null coordinates. 
% } %%% 
%
%As shown in~\cite{HI05}, supertranslation does not exist 
%in higher dimensions. This can be manifest and easy to compare 
%with the $4$-dimensional case when one examines asymptotic translational 
%symmetries in terms of the Gaussian null coordinates. 
% 
Then, we provide a (modified) definition of the news tensor, which is 
slightly different from that given in \cite{HI05} and is now regular 
under the Gaussian null conformal gauge. 
We give our main theorem that indicates the existence of 
a Bondi energy-momentum vector in arbitrary even dimensions greater 
than or equal to $4$, and then derive 
an expression of a higher dimensional Bondi-energy in terms of 
our conjectured Bondi energy-momentum vector. 
The energy-loss formula is also given there. 
Summary and brief discussion on the regularity of our 
Bondi energy-momentum vector are given in Section~\ref{sect:discussion}. 
The conventions of the metric signature and definitions of 
curvature tensors follow~\cite{Wald84}. The spacetime dimension, 
denoted by $d$, is assumed to be even throughout the paper, 
unless otherwise stated.  	

\section{Asymptotic flatness in higher (even) dimensions}  
%---------------------------------------------------------------------------

In this section, we shall recapitulate the results of the paper \cite{HI05} 
(see also \cite{HI05review}). We first state our boundary conditions 
that define asymptotic flatness at null infinity in higher dimensions 
and then review a general strategy for defining conserved quantities 
in asymptotically flat spacetimes. 
After deriving the flux formula and the Bondi-energy 
in higher dimensions under the Minkowskian conformal gauge, 
we see why that gauge specifies non-compact cross-sections 
of null infinity. 

\subsection{Stable definition of asymptotic flatness}    
%---------------------------------------------------------------------------

Let $(M,g_{ab})$ be a physical spacetime of dimension $d$ 
which we wish to identify as an asymptotically flat spacetime 
in a certain sense. 
We are interested in, within the framework of conformal infinity, 
the question of how to specify the precise rate at which Minkowski spacetime, 
$(M,\eta_{ab})$, is approached $(M,g_{ab})$, at large distances 
toward null infinity. For this purpose it is convenient to introduce 
the following two fictitious spacetimes. 
% \medskip 
Let $({\tilde M}, {\tilde g}_{ab})$ and $({\bar M}, {\bar g}_{ab})$ be 
an {\sl unphysical conformal spacetime} and the {\sl background geometry} 
associated with $(M,g_{ab})$. 
These two metrics, $\tg_{ab}$ and ${\bar g}_{ab}$, 
are related to the physical metric, $g_{ab}$, and the Minkowski metric, 
$\eta_{ab}$, via a smooth {\sl conformal factor}, $\Omega$, as 
${\tilde g}_{ab} = \Omega^2 g_{ab}$ and ${\bar g}_{ab} = \Omega^2 \eta_{ab}$ 
so that one can attach a {\sl boundary} $\I$ at $\Omega=0$ to $M$ such that 
there exists an open neighborhood of $\I$ in ${\tilde M} = M \cup \I$ 
which is diffeomorphic to an open subset of the manifold $\bar M$,  
and $\I$ is mapped to a subset of the boundary of ${\bar M}$. 
(See \cite{HI05} for more details.)
As usual, indices on tensor fields on $\tilde M$ with ``tilde'' are 
lowered and raised with the unphysical metric, $\tg_{ab}$, and its 
inverse, and a similar rule applies to tensor fields 
with ``bar'' on $\bar M$. Note that $\I$ is divided into disjoint sets, 
$\I^+$ and $\I^-$, the future and past null infinity, respectively. 
In the following, by $\I$ we mean the future null infinity $\I^+$, 
unless otherwise stated. 
We assume that $\I$ be topologically ${\Bbb R} \times S^{d-2}$  
so that it is consistent with the notion of a (higher dimensional version 
of) {\sl weakly asymptotically simple} spacetime. 
% (See e.g., \cite{Wald84}.)
% the topology of the boundary of our 
% background geometry $(\bar M, \bar g_{ab})$.    

\medskip 
We call $(M,g_{ab})$ of $d$ (even) dimensions 
{\sl asymptotically flat at null infinity} if the corresponding unphysical 
metric satisfies the following boundary (fall off) conditions near null 
infinity $\I$,  
\bena 
{\tilde g}_{ab} - {\bar g}_{ab} = O(\Omega^{(d-2)/2}) \,, 
\quad 
&& 
 {\tilde \epsilon}_{ab\cdots c} - {\bar \epsilon}_{ab\cdots c} 
 = O(\Omega^{d/2}) \,, 
\non \\ 
\left( {\tilde g}^{ab} - {\bar g}^{ab} \right) (\dd \Omega)_b
= O(\Omega^{d/2}) \,, 
\quad 
&& 
\left( {\tilde g}^{ab} - {\bar g}^{ab} \right) (\dd \Omega)_a(\dd \Omega)_b
 = O(\Omega^{(d+2)/2}) \,,  
\label{condi:asympt-flat}
\eena
where $\tilde \epsilon$ and $\bar \epsilon$ denote, respectively, 
the natural volume element of the unphysical spacetime and 
the background geometry. 

\bigskip 
The definition of asymptotic flatness above is arrived at through 
an analysis of stability of conformal null infinity against linear 
gravitational perturbations. 
% \paragraph{Theorem 1:} 
More precisely, let $(M,g_{ab})$ be a globally hyperbolic solution 
to the vacuum Einstein equations and $\delta g_{ab}$ be a solution to 
the linearized vacuum Einstein equations with initial data of compact 
support on a Cauchy surface. Then, it is shown \cite{HI05} that 
there exists a gauge for all even $d>4$ so that the unphysical 
metric perturbations, $\delta {\tilde g}_{ab} = \Omega^2 \delta g_{ab}$, 
behave at $\I$ as 
\bena 
 \delta {\tilde g}_{ab} = O(\Omega^{(d-2)/2}) \,, \quad 
 \delta {\tilde g}_{ab} \tn^b = O(\Omega^{d/2}) \,, \quad 
 \delta {\tilde g}_{ab} \tn^a \tn^b = O(\Omega^{(d+2)/2}) \,, \quad 
 {\tilde g}^{ab}\delta {\tilde g}_{ab} = O(\Omega^{d/2}) \,,  
\label{condi:af:linear}
\eena  
where $\tn^a = \tg^{ab}\tnabla_b \Omega$. 

\bigskip 
One can view eqs.~(\ref{condi:af:linear}) as a linearized version of 
the asymptotic flatness conditions above, eqs.~(\ref{condi:asympt-flat}). 
In $4$-dimensions, linear stability of conformal null infinity was 
shown by Geroch and Xanthopoulos \cite{GX78}, in which 
the corresponding drop-off conditions are given by 
\bena
 \delta {\tilde g}_{ab} = O(\Omega) \,, \quad 
 \delta {\tilde g}_{ab} \tn^b = O(\Omega^{2}) \,, \quad 
 \delta {\tilde g}_{ab} \tn^a \tn^b = O(\Omega^{3}) \,, \quad 
 {\tilde g}^{ab}\delta {\tilde g}_{ab} = O(\Omega) \,.   
\label{condi:af:linear:4d}
\eena
We note that in $4$-dimensions, eqs.~(\ref{condi:af:linear:4d}), 
the trace-part of the perturbation falls off as fast as the other components 
of the metric perturbation, in contrast to the case of higher $d>4$ 
dimensions, eqs.~(\ref{condi:af:linear}).    
We also should note that the transverse-traceless gauge works 
when $d>4$, but does not when $d=4$; one needs to use 
Geroch-Xanthopoulos gauge \cite{GX78} for $d=4$.  
Under these gauge conditions, one can show that 
the vacuum Einstein equations with certain field variables that correspond 
to relevant unphysical metric perturbations---such as 
$\Omega^{-(d-2)/2} \delta {\tilde g}_{ab}$---indeed form 
a hyperbolic system of partial differential equations that possesses 
a well-posed initial value formulation in the unphysical spacetime 
and thereby conclude that for compactly supported smooth initial data 
for perturbations, the field variables extend to smooth tensor fields 
at null infinity, $\I$. 

\medskip 
It should be commented that the unphysical metric generically fails to be 
smooth at null infinity in odd-spacetime dimensions 
when gravitational radiation presents; 
it follows from half-integral powers of $\Omega$ in 
eqs.~(\ref{condi:af:linear}) that the unphysical metric appears to 
be at most $(d-3)/2$ times differentiable at null infinity.  
This non-smoothness of the unphysical metric in odd dimensions 
% at conformal null infinity in odd-dimensional radiative spacetime 
has been shown more explicitly in 
\cite{Hollands-Wald04} by examining the leading order behavior of 
the unphysical Weyl curvature perturbations off of Minkowski spacetime. 
Mainly for this reason, our conformal approach to defining 
stable notion of asymptotic flatness at null infinity does not 
apply to odd dimensional spacetimes. 

\medskip 
Note also that as mentioned above, the asymptotic flatness (fall-off) 
conditions, eq.~(\ref{condi:asympt-flat}), are not derived from analysis of 
full non-linear theory but rather postulated based on linear stability 
analysis of conformal null infinity against gravitational perturbations.  
Higher order perturbation effects---if taken into account---might possibly 
alter the asymptotic fall-off behavior, but in the present paper, we 
are not going to investigate such non-linear effects on the asymptotic 
flatness conditions.  
% Examining such non-linear effects on the asymptotic flatness 
% conditions is, however, beyond the scope of the present paper. 

\medskip 
Once boundary conditions for asymptotic flatness at null infinity are 
established, one can introduce the notion of asymptotic symmetries. 
Consider vector fields $\xi^a$ on $\tilde M$ which are smooth, complete 
on $\tilde M$, and in particular tangent to $\I$ on $\I$, and are such that 
\ben
\Omega^2 \pounds_\xi g_{ab} = \pounds_\xi {\tilde g}_{ab} 
  - 2 \Omega^{-1}\xi^c \tn_c \tg_{ab} 
\een 
satisfies the (linearised version of) asymptotic flatness conditions, 
eq.~(\ref{condi:af:linear}) for $d>4$ and 
eq.~(\ref{condi:af:linear:4d}) for $d=4$. 
Since we are concerned with symmetry properties at null infinity $\I$, 
we introduce an equivalence relation into the set of vector fields 
$\xi^a$ by viewing them as {\sl equivalent} if they coincide on $\I$. 
An equivalence class of such vector fields---we denote it by 
the same notation $\xi^a$ in the following---is called 
an {\sl infinitesimal asymptotic symmetry} and generates 
an {\sl asymptotic symmetry group}, which is a one-parameter group 
of diffeomorphisms $\phi$ which map any asymptotic flat metric $g_{ab}$ 
to an asymptotic flat metric ${\phi^* g}_{ab}$.

\subsection{Strategy for defining Bondi energy} 
%---------------------------------------------------------------------------

Based on what is sometimes called the {\sl covariant phase space} method, 
Wald and Zoupas \cite{Wald-Zoupas00} developed a general strategy 
for defining conserved quantities associated with symmetries that preserve 
a given set of boundary conditions, which can apply to any theories derived 
from a diffeomorphism covariant Lagrangian. 
%
% The strategy is sometime referred to as covariant phase space method.  
(For earlier work of the covariant phase space method, see 
e.g., \cite{CW87,LeeWald90,ABL91} and references therein.) 
For Einstein's gravity, 
our starting point is the Lagrangian density $d$-form, $L$, given 
in terms of the scalar curvature, $R$, and the natural $d$-volume element, 
$\epsilon$, with respect to the physical metric, $g_{ab}$, as 
\ben
 L = \frac{1}{16\pi G} R \epsilon \,, 
\een 
with the boundary null infinity $\I$ and the boundary conditions 
specified in the definition of asymptotic flatness above, 
eqs.~(\ref{condi:af:linear}) and (\ref{condi:af:linear:4d}), 
where associated are asymptotic symmetries $\xi^a$. 
Taking variation with respect to the metric $g_{ab}$, one has 
\ben
  \delta L = E + \dd \theta \,, 
\een
where $E$ denotes Einstein's equations and $\theta$ is a $(d-1)$-form given by 
\ben
\theta_{a_1 \dots a_{d-1}}  = \frac{1}{16\pi G} 
g^{ab}g^{cd}(\nabla_a\delta g_{bd} - \nabla_d \delta g_{ab}) 
\epsilon_{ca_1\dots a_{d-1}} \,.  
\een 
Consider a pair of variations $(\delta_1 g,\, \delta_2 g)$ and 
assume that the two variations commute, i.e., $\delta_1 \delta_2 g - 
\delta_2 \delta_1 g = 0$. 
Then, the symplectic current, $\omega$, can be defined as 
\ben
  \omega (g; \,\delta_1 g, \, \delta_2 g) = 
 \delta_1 \theta (g;\delta_2 g) - \delta_2 \theta (g;\delta_1 g) \,.
\een 
If $\omega$ has a {\sl vanishing} extension to the boundary 
under consideration, 
then there exists a conserved Hamiltonian associated with $\xi^a$. 
However, this is, in general, not the case especially 
when radiation presents near the boundary $\I$. However, even in that case, 
if (i) $\omega$ has a well-defined (finite) extension to $\I$ for any 
asymptotically flat metric and in addition (ii) there exists a symplectic 
potential, $\Theta$, on $\I$ such that 
\ben
 \zeta^* \omega (g; \, \delta_1 g, \, \delta_2 g) 
 = \delta_1 \Theta(g;\delta_2 g) - \delta_2 \Theta(g;\delta_1 g) \,, 
\label{zeta:omega}
\een
where here and in the following $\zeta^*$ denotes the pull back of 
a tensor field to $\I$, then one can define an associated charge 
${\cal H}_\xi$ by 
\ben
 \delta {\cal H}_\xi = \int_B \left(\delta Q[\xi] - \xi \cdot \theta \right)
 + \int_B \xi \cdot \Theta \,, 
\label{def:charge-Hxi}
\een
where ``$\cdot$'' denotes the interior product, $B$ is a given cross section 
at $\I$, and where 
\ben
  Q[\xi]{}_{a_1 \cdots a_{d-2}} 
  = - \frac{1}{16\pi G} (\nabla^b \xi^c) \epsilon_{a_1 \cdots a_{d-2}bc} 
\label{def:Noethercharge}
\een 
is the Noether charge $(d-2)$-form. 
Note that the integrand, $\xi \cdot \Theta$, of the second integral 
of the right-hand side of eq.~(\ref{def:charge-Hxi}) is, by definition, 
a smooth quantity on $\I$, whereas the integrand, 
$\delta Q[\xi] - \xi \cdot \theta$, of the first integral is 
defined only inside the spacetime. 
Therefore, 
here and in the following, 
an integral over a given cross-section $B$ of $\I$ should be understood 
as follows: Consider first an achronal hypersurface $\Sigma$ in some 
neighborhood of $\I$ which extends to $B$ and a nested sequence of 
compact subsets $K_i$ of $\Sigma$ such that 
as $i \rightarrow \infty$, the $(d-2)$-surface $S_i \equiv \partial K_i$ 
approaches $\partial \Sigma = B$. Then, evaluate the integral over $S_i$ and 
then take the limit $i \rightarrow \infty$. 
The existence of the limit follows from the same argument below eq.~(45) 
in \cite{HI05} when $\omega$ has a smooth extension to $\I$.  
% 
% \footnote{%%% 
% The integrand, $\Theta$, of the second integral of the right-hand side 
% of eq.~(\ref{def:charge-Hxi}) is, by definition, a smooth quantity 
% on $\I$, whereas the integrand, $\delta Q[\xi] - \xi \cdot \theta$, 
% of the first integral is defined only inside the spacetime. 
% For this reason, we need to define an integral over $B$ on $\I$ as 
% the limit to $B$ of the corresponding integrals defined over closed 
% $(d-2)$-surfaces inside the spacetime. 
% } %%% 
% 
%
Note also that field variations taken in eq.~(\ref{def:charge-Hxi}) are 
assumed to satisfy the linearized Einstein equations.

\medskip 
In the $4$-dimensional case, it was shown \cite{Wald-Zoupas00} that 
under certain fall-off conditions and the choice of the Bondi gauge, 
the symplectic current $3$-form, $\omega$, extends smoothly to $\I$ and, 
in general, non-vanishing there (hence the assumptions (i) and (ii) hold)  
and the resultant ${\cal H}_\xi$ indeed agrees with the Bondi energy 
expression \cite{Geroch77}.

\medskip 
The assumptions (i) and (ii) were shown to hold also 
in higher (even) dimensions~\cite{HI05}, under the boundary conditions, 
eq.~(\ref{condi:asympt-flat}). 
Therefore, according to Wald and Zoupas~\cite{Wald-Zoupas00}, one is able 
to construct a charge ${\cal H}_\xi$ associated with an asymptotic symmetry 
$\xi^a$ also in higher dimensions. 
% satisfying the defining equation~(\ref{def:charge-Hxi}).  
The proposal of \cite{HI05} is that one can take ${\cal H}_\xi$ 
as the definition of a higher dimensional Bondi energy. 
For this purpose, we seek for a vector field $P^a$ on $\tilde M$ 
that satisfies $P^a \tn_a = O(\Omega^2)$ and also 
\ben
{\Theta} 
 = {}^{(d-1)}\!{\tilde \epsilon} \; {\tilde \nabla}_a P^a 
   + O(\Omega) \,,  
\label{def:eq:P}  
\een 
where 
${}^{(d-1)}\!{\tilde \epsilon}$ is the volume element on $\I$ 
defined by ${\tilde \epsilon}_{a_1\cdots a_d} = d \tn_{[a_1} 
{}^{(d-1)}\!{\tilde \epsilon}_{a_2\cdots a_d]}$.  
Note that above eq.~(\ref{def:eq:P}) determines $P^a$ uniquely, 
up to the addition of a vector field $X^a$ which satisfies 
$X^a \tn_a = O(\Omega^2)$ and 
$
{}^{(d-1)}\!{\tilde \epsilon} \: \tnabla_aX^a = \dd \mu + O(\Omega)
% \label{eq:freedom-of-P}
$
with $\mu$ being some $(d-2)$-form. In other words, if $P^a$ is a solution 
to eq.~(\ref{def:eq:P}), then 
\ben 
P'^a \equiv P^a + \tnabla_b X^{ab} 
\label{eq:freedom-of-P-by-X} 
\een 
also is a solution to eq.~(\ref{def:eq:P}), provided that $X^{ab}$ is 
an anti-symmetric tensor on $\tilde M$ such that 
$ \tn_a \tnabla_b X^{ab}= O(\Omega^2)$ in a neighborhood of $\I$. 
%
% For such a solution $P^a$, one can then get the formula, 
% \ben  
% % \tnabla_{[{a_1}}({\tilde \epsilon}_{ {a_2} \cdots {a_{d-1}}] bc} 
% % \ell^b P^c)
% \dd \mu = {}^{(d-1)}\!{\tilde \epsilon} \; {\tilde \nabla}_a P^a  
% + \dd \Omega \wedge \varphi
% %-2(d-1) \tnabla_b(P^{[b}\ell^{c]}) n_{[{a_1}}
% %                             {\tilde \epsilon}_{|c|{a_2}\cdots {a_{d-1}}]}
% + O(\Omega) \,,  
% \label{formula:Stokes}
% \een 
% where $\mu_{{ {a_1} \cdots {a_{d-2}}}} = 
% {\tilde \epsilon}_{ {a_1} \cdots {a_{d-2}} bc} \tell^b P^c$ and 
% $\varphi_{ {a_1} \cdots {a_{d-2}} } =2\tnabla_b(P^{[b}\tell^{c]})
% {\tilde \epsilon}_{c{a_1}\cdots {a_{d-2}}}$ with 
%$\tell^a$ being a null geodesic tangent in $\tilde M$ 
%that satisfies $\tell^a\tnabla_a \Omega =1$, so that $\tell^a$ 
%is transverse to $\I$. 
Now let $I \subset \I$ be a segment of $\I$ bounded by two cross-sections,  
$B$ and $B_0$ (with $B$ being in the future of $B_0$).  
Integrating the above formula, eq.~(\ref{def:eq:P}), 
over $I \subset \I$ and applying Stokes' theorem, 
one formally obtains the following formula,  
\bena
 F_\xi \equiv % {\cal H}_\xi (B_2) - {\cal H}_\xi (B_1) 
        - \int_{I} {\Theta}(g;\pounds_\xi g) 
        = \int_{B} {}^{(d-2)}\!{\tilde \epsilon} \:P^a \tell_a 
        - \int_{B_0} {}^{(d-2)}\!{\tilde \epsilon} \:P^a \tell_a  \,,     
\label{def:flux}
\eena
where ${}^{(d-2)}\!{\tilde \epsilon}$ denotes the natural volume element 
on a cross-section $B$ induced from $\tilde \epsilon$, and where here 
and hereafter $\tell^a$ denotes a null geodesic tangent in $\tilde M$ 
that satisfies $\tell^a\tnabla_a \Omega =1$, so that $\tell^a$ 
is transverse to $\I$.  
Note that $F_\xi$---which corresponds to the flux through the segment 
$I$ associated with an infinitesimal symmetry $\xi^a$---is well-defined, 
and finite since by definition $\Theta$ itself is a smooth, finite 
quantity on $\I$ under our boundary conditions. 
Thus, if each integral of the right-hand side of eq.~(\ref{def:flux}) is 
well-defined by itself (and independent of how $S_i$ approaches $B$), 
one can identify the (higher dimensional) Bondi energy at $B$ with 
\bena
  {\cal H}_\xi = \int_B {}^{(d-2)}\!{\tilde \epsilon} \:P^a \tell_a \,,   
\label{def:H}
\eena
and may view $P^a$ as a (higher-dimensional) Bondi energy-momentum integrand.  
% This is indeed the case in $4$-dimensions. 

\subsection{Bondi energy in Minkowskian conformal gauge} 
%---------------------------------------------------------------------------

Now we construct the symplectic potential, $\Theta$, 
for eq.~(\ref{zeta:omega}). For this purpose, 
let us introduce the following tensor field 
\ben  
 {\tilde S}_{ab} = 
 \frac{2}{d-2} {\tilde R}_{ab} - \frac{1}{(d-1)(d-2)} 
                                 {\tilde R} {\tilde g}_{ab} \,,          
\label{eq:vac:Einstein}
\een  
where ${\tilde R}_{ab}$ and ${\tilde R}$ are the Ricci tensor 
and the scalar curvature with respect to $\tg_{ag}$.  
Using the standard formulas for conformal transformation of a metric 
and associated curvature tensors (see e.g., \cite{Wald84}), 
one can rewrite the vacuum Einstein's equation, $R_{ab}=0$, as 
\ben
 {\tilde S}_{ab} = - 2{\Omega}^{-1}{ {\tilde \nabla}_{(a} n_{b)} }
                  + {\Omega^{-2}}{{\tilde n}^c n_c}{\tilde g}_{ab} \,. 
\label{eq:conf-Einstein}
\een
Then, the pull back to $\I$ of the symplectic current $(d-1)$-form, $\omega$, 
can be expressed as 
\bena
 \zeta^* \omega_{a_1 \cdots a_{d-1}} 
 = - \frac{1}{32\pi G} \Omega^{-(d-3)} 
    \left( 
           \delta_2 {\tilde g}^{bc}\delta_1 {\tilde S}_{bc}  
         - \delta_1 {\tilde g}^{bc}\delta_2 {\tilde S}_{bc} 
    \right) {\tilde \epsilon}_{a_1 \cdots a_{d-1}} \,.   
\label{symp:curr:S} 
\eena 
Note that from the asymptotic flatness conditions, eq.~(\ref{condi:af:linear}),
it follows that 
\bena
 {\tilde S}_{ab} - {\bar S}_{ab} = O(\Omega^{(d-4)/2}) \,, \quad 
 {\tilde S}^c{}_c - {\bar S}^c{}_c = O(\Omega^{(d-2)/2}) \,, \quad 
 {\tilde n}^a{\tilde n}_a - {\bar n}^a  {\bar n}_a = O(\Omega^{(d+2)/2}) \,. 
\eena
At this point, one can introduce a trace-free symmetric tensor, 
$N_{ab}$, defined on $\I$ so that $\zeta^* \omega$ above can be expressed as 
\ben
 \zeta^* \omega_{a_1 \cdots a_{d-1}} 
 = - \frac{1}{32\pi G} \Omega^{-(d-2)/2} 
    \left( 
           \delta_2 {\tilde g}^{bc}\delta_1 N_{bc}  
         - \delta_1 {\tilde g}^{bc}\delta_2 N_{bc} 
    \right) {\tilde \epsilon}_{a_1 \cdots a_{d-1}} \,.   
\label{symp:curr:N} 
\een 
The symplectic potential is then expressed as 
\ben
  {\Theta}({g}; \delta {g})_{a_1 \cdots a_{d-1}} 
  = \frac{-1}{32\pi G} \Omega^{-(d-2)/2} \delta {\tilde g}^{bc}N_{bc} \: 
    {\tilde \epsilon}_{a_1 \cdots a_{d-1}} \,.  
\label{SymPote:Minkowskigauge}
\een

\medskip 
In order to define $N_{ab}$---called the {\sl news tensor}---in terms 
of ${\tilde S}_{ab}$ and obtain an explicit formula for the Bondi-energy, 
we impose gauge conditions. Perhaps the simplest gauge choice is 
to impose 
\ben
 {\bar S}_{ab} = 0 \,, \quad {\bar n}^a{\bar n}_a = 0 \,, \quad 
{\bar \nabla}_a {\bar n}^b = 0 \,.   
\label{gauge:flat}
\een  
This gauge choice can be taken at least locally in some neighborhood of $\I$, 
where ${\bar n}^a = {\bar g}^{ab} {\bar \nabla}_b \Omega$.  
This gauge---the {\sl Minkowskian conformal gauge}---employed 
in \cite{HI05} corresponds to requiring that 
the background metric, ${\bar g}_{ab}$, be the flat metric 
in a neighborhood of the boundary of $\bar M$, 
and the quantities on the unphysical manifold ${\tilde M}$ satisfy 
\ben
 {\tilde S}_{ab} = O(\Omega^{(d-4)/2}) \,, \quad 
 {\tilde n}^a {\tilde n}_a = O(\Omega^{(d+2)/2}) \,, \quad 
 {\tilde \nabla}_a {\tilde n}_b = O(\Omega^{(d-2)/2}) \,.   
\label{condi:flatgauge}
\een 
The news tensor is then defined by (eq.~(61) in \cite{HI05}) 
\ben
 N_{ab} = \zeta^* (\Omega^{-(d-4)/2} q^m{}_a q^m{}_b \tilde{S}_{mn}) \,, 
\label{def:news:flat}
\een
where $q_{ab} = {\tilde g}_{ab} - 2 \tell_{(a}n_{b)}$, 
with $\tell_a$ being a covector on ${\tilde M}$ such that 
${\tilde \nabla}_a \tell_b = O(\Omega^{(d-4)/2})$, 
${\tilde \ell}^a \tell_a = O(\Omega^{(d-2)/2})$, 
${\tilde n}^a \tell_a = 1 + O(\Omega^{d/2})$ near $\I$. 
Under the Minkowskian conformal gauge, 
expressing $\delta \tg_{ab}= \Omega^2 \pounds_\xi g_{ab}$ in terms of 
${\tilde S}_{ab}$ via the Einstein equations, eq.~(\ref{eq:conf-Einstein}),  
and then substituting into eq.~(\ref{SymPote:Minkowskigauge}), 
one obtains, for the translation $\xi^a = \tau \tn^a$ with $\tau = const.$,  
\ben
 \Theta (g; \pounds_{\xi} g) 
 = \frac{1}{32 \pi G}\tau \Omega^{-(d-4)} S_{ab}S_{cd}q^{ac}q^{bd} \,. 
\een
The Bondi energy-momentum, $P^a$, for $\xi^a = \tau \tn^a$ is then identified 
with 
\ben 
  P^a = \frac{\tau}{8 (d-3)\pi G} \Omega^{-(d-4)} 
      \left(
        {\tilde n}^{[a}{\tilde q}^{b]d}{\tilde q}^{ce}{\tilde S}_{de}
        {\tilde \nabla}_b \tell_c 
      - \Omega^{-1} {\tilde C}^{abcd} n_b \tell_c n_d 
      \right) \,,  
\label{def:P:Minkowskiangauge}
\een
and the Bondi-energy for $\xi^a = \tau \tn^a$ is given by \cite{HI05}
\bena 
 {\cal H}_\xi 
  &=& \frac{1}{8(d-3) \pi G} \int_B \tau \Omega^{-(d-4)} 
   \left\{  
        \frac{1}{(d-2)} {\tilde R}_{ab}q^{ac}q^{bd}({\tilde \nabla}_c \tell_d) 
        - \Omega^{-1} {\tilde C}^{abcd} n_b \tell_c n_d 
   \right\}\: {\tilde \epsilon} \,,  
\eena 
where ${\tilde C}_{abcd}$ denotes the Weyl tensor with respect to $\tg_{ab}$.  

\medskip 
The Minkowskian conformal gauge can be achieved at least locally 
but does not seem to globally determine background structure 
with desired property. 
As a simple case, let us consider Minkowski spacetime $(M,\eta_{ab})$ 
as our physical spacetime. Then, the Minkowskian conformal gauge, 
eq.~(\ref{gauge:flat}), 
% under which the background geometry 
% $({\bar M}, {\bar g}_{ab})$ becomes flat 
can be explicitly constructed as follows. 
% Then, the corresponding unphysical spacetime 
% under the Minkowskian conformal gauge, eq.~(\ref{gauge:flat}), can be 
% explicitly constructed by considering a map from a portion of $\I$ of 
% Minkowski spacetime to a Rindler horizon. More concretely, 
Let $x^\mu = (x^0,x^1, x^i)$, $i= 2 \cdots (d-1)$ be Cartesian coordinates  
in $(M,\eta_{ab})$, and let $V^+ = \{x | x_\mu x^\mu <0,\:x^0>0\}$ 
i.e., interior of the future light cone of the origin $x^\mu =0$, 
and $W = \{ x | x^1 \geq |x^0|\}$, a Rindler wedge. Consider the map $\phi$ 
on $(M,\eta_{ab})$ defined by (see eq.~(B8) in \cite{HI05}), 
\ben
 \phi^\mu(x) = 
        \frac{a^\mu + b^\mu x_\nu x^\nu 
            + 2\left(\eta^{\mu \nu} + a^{(\mu} b^{\nu)}\right)x_\nu
             }{2b_\sigma x^\sigma} \,, 
\een 
where $a^\mu = (1,-1,0,\cdots,0)$ and $b^\mu = (1,1,0,\cdots,0)$ in   
the above coordinates. 
One can observe that $\phi$ maps points in $V^+$ bijectively to 
points in $W$ and furthermore 
\ben
(b_\sigma x^\sigma)^2 \eta_{\mu \nu} \dd \phi^\mu \dd \phi^\nu 
 = \eta_{\mu \nu} \dd x^\mu \dd x^\nu \,.  
\een 
Thus, $\phi$ is a conformal isometry from $V^+$ to $W$. 
Therefore if we choose $\bar{M}=W$ and 
$\bar{g}_{ab} = \Omega^2 \phi_*\eta_{ab}$ with $\Omega = b_\mu x^\mu$, 
then our background geometry $(\bar{M},\bar{g}_{ab})$ satisfies 
the Minkowskian conformal gauge conditions.  
Since defining a Rindler wedge $W$ needs to choose a particular spatial 
direction---the direction specified by the vector $b^\mu$ 
in the example above, the conformal isometry $\phi:V^+ \rightarrow W$ 
breaks the {\sl global isotropy} the original Minkowski spacetime possesses.  
As a result, all null generators of $\I$ are mapped to $\partial W$, 
except a single null generator of $\I$ in the direction specified by $b^\mu$.  
Consequently, a cross-section $B$ on $\partial \bar{M}$ specified 
by this gauge choice is a $(d-2)$-sphere with a single point removed, 
and hence non-compact. 
Although the above example of the Minkowskian conformal gauge is given 
in the special context that $(M,g_{ab})$ is Minkowski spacetime, 
it seems highly unlikely that the Minkowskian conformal gauge would 
admit a compact cross-section in $\I$ for more generic, asymptotically 
flat spacetimes.

\section{Gaussian null conformal gauge} 
\label{sect:GNC} 
%---------------------------------------------------------------------------

In this section, instead of the Minkowskian conformal gauge discussed above, 
we shall consider Gaussian null conformal gauge, which allows us to take 
globally defined, compact spherical cross-sections of $\I$.  
We shall discuss the asymptotic flatness conditions, definition of 
news tensor, and the Bondi energy-momentum under the Gaussian null 
conformal gauge condition.  

\subsection{Flatness conditions and asymptotic symmetries}
%---------------------------------------------------------------------------

As seen in the previous section, in even spacetime dimensions, a conformal 
null infinity, $\I$, can always exist as a smooth null hypersurface 
in an unphysical spacetime $(\tilde M,\,\tilde g_{ab})$, being stable 
against at least linear perturbations. 
Then, following the standard procedure (Appendix A of \cite{FRW99}), 
one can construct a coordinate system $x^\mu = (u,\Omega,x^A)$ 
with $A=1,\dots, d-2$ on a (sufficiently small open) neighborhood, $\cal O$, 
of an arbitrary point $p\in \I$ in $\tilde M$ such that in $\cal O$, 
the unphysical metric takes the form, 
%for any point $p$ on a null generator $\gamma$ of $\I$, 
%one can always find a neighborhood $U_p \subset \tilde M$ and coordinates 
%$x^\mu = (u,\Omega,x^A)$ with $A=1,\dots, d-2$ such that in $U_p$, 
%the unphysical metric takes the form (see e.g., \cite{MI83,Racz07}) 
\ben
 \dd \tilde s^2 = \tg_{\mu \nu}\dd x^\mu \dd x^\nu 
                =\tilde \alpha \dd u^2 + 2 \dd u \dd \Omega 
                  + 2\tilde \beta_A \dd u \dd x^A 
                  + \tilde \gamma_{AB} \dd x^A \dd x^B \,,    
\label{def:gnc}  
\een
where $u$ parametrizes a congruence of null generators of $\I \cap {\cal O} $ 
with $(\partial/\partial u)^a$ being a tangent vector field of 
the congruence, and where $\Omega$, chosen to be $\Omega=0$ on 
$\I \cap {\cal O} $, is an affine parameter of null geodesics which are 
orthogonal to each $(u,\Omega)=const.$ surfaces $B(u,\Omega)$ in $\cal O$ 
and transverse to $\I \cap {\cal O} $ so that 
${\tilde g}_{ab}(\partial/\partial \Omega)^a(\partial/\partial u)^b = 1$, 
and where $\tilde \alpha$, $\tilde \beta_A$, $\tilde \gamma_{AB}$ 
are smooth functions with $\tilde \alpha =0 = \tilde \beta_A$ on 
$\I \cap {\cal O}$. 
% of $B(u,\Omega)$ (at least when $\Omega =0$). 
% 
Note that $x^A= (x^1,\dots,x^{d-2})$ may be regarded as local coordinates 
on $B(u,\Omega)$ and also that $\tilde \gamma_{AB} \dd x^A \dd x^B$ 
is a Riemaniann $(d-2)$-metric, which does not necessarily coincide 
with the induced metric on $B(u,\Omega)$ when $\Omega \neq 0$.    
%%% 
The chart, $({\cal O},x^\mu)$, introduced in this way is called the 
{\sl Gaussian null coordinate system} with respect to the null surface 
$\I \cap {\cal O}$ because of its similarity to Gaussian {\sl normal} 
coordinates with respect to a timelike or spacelike hypersurface in 
a spacetime. 
% \footnote{ %%% 
%A construction of such a coordinate system has been discussed in, 
%for example, \cite{MI83}, in which the relevant null hypersurfaces 
%are ruled by closed null geodesics and consequently are diffeomorphic 
%to $B\times S^1$ with some $2$-manifold $B$. 
%However, there seems to be no obstruction in generalizing the arguments 
%of coordinates conditions (the arguments around eq.~(2.2), or 
%eqs.~(2.5) and (2.6) in \cite{MI83}) to the construction of Gaussian null 
%coordinates with respect to non-compact null hypersurface 
%$B \times {\Bbb R}$ with some $(d-2)$-manifold~$B$. 
% } %%% 

%%% 
It is, in general, not obvious when one can construct a Gaussian null 
coordinate system that covers the entire $\I$ so that, in particular, 
the set of $B(u)=B(u,\Omega =0)$ becomes a {\sl global} foliation of $\I$; 
each $B(u)$ is, in general, an open subset of global cross-sections 
of $\I$ and one may need to patch together more than one coordinate chart 
to cover $\I$. 
%%% 
However, when, for example, a null hypersurface is ruled by some Killing 
vector field generating a one-parameter group of isometries, 
one can construct, by patching together local results, essentially a global 
Gaussian null coordinate system that covers the entire null hypersurface 
(see e.g., \cite{HIW07}). 
Although asymptotically flat spacetimes we are concerned with here 
do not necessarily have a Killing symmetry, they do admit asymptotic 
symmetries, $\xi^a$, which are tangent to $\I$ and play a similar role 
of a Killing symmetry on $\I$.   
%(and in particular, one of which is normal to $\I$, corresponding 
% to time-translation), 
Therefore there seems to be no obstruction in assuming that one always be 
able to construct a desired, global Gaussian null coordinate system 
%$({\cal O},x^\mu)$ 
in some neighborhood $\cal O$ of $\I$ in $\tilde M$ such that 
% in even dimensional asymptotically flat spacetimes 
$B(u)$ appropriately foliate $\I$ as global cross-sections of 
$\I$ with topology $B(u) \approx S^{d-2}$, reflecting 
$\I \approx {\Bbb R} \times S^{d-2}$.  
In the following, we impose the existence of a global Gaussian 
null coordinate system, $({\cal O},x^\mu)$, with respect to $\I$, as a part 
of our conditions for asymptotic flatness at null infinity.
\footnote{ %%%
We believe that a global Gaussian null coordinate system with respect to 
$\I$ can always be constructed under the assumptions that have been made 
in the present paper, but we have not fully investigated this issue. 
A formal proof for the global existence of such a desired coordinate 
system needs to be given.   
} %%%  

\medskip 
For our background geometry, $({\bar M}, \bar{g}_{ab})$, the 
Gaussian null chart $({\cal O}, x^\mu)$ yields 
\ben
{\bar \alpha} = -\Omega^2 \,, \quad  {\bar \beta}_A =  0 \,, \quad 
{\bar \gamma}_{AB} = \sigma_{AB} \,,    
\label{def:bckgrd:gnc}
\een  
and globally covers $\partial {\bar M}$ (which is diffeomorphic to $\I$), 
where here and hereafter $\sigma_{AB}$ denotes the metric of 
$(d-2)$-dimensional unit round sphere.  
We shall view $\sigma_{AB}$ as a {\sl global specification of cross sections 
of $\I$} by the choice of gauge, eq.~(\ref{def:gnc}). 
In the following we call the gauge choice, eqs.~(\ref{def:gnc}) and  
(\ref{def:bckgrd:gnc}), the {\sl Gaussian null conformal gauge}. 

\medskip 
Note that viewing $\Omega$ as a conformal factor and relating it to 
the {\sl luminosity distance}, $r$, by $r= 1/\Omega$, 
one can find the associated {\sl physical} Gaussian null coordinate system,  
$(u,r,x^A)$, with which the physical metric is written 
\ben 
 r^2 \dd {\tilde s}^2 = \dd s^2 
           = \alpha \dd u^2 - 2 \dd u \dd r 
           + 2 \beta_A \dd u \dd x^A 
           + \gamma_{AB} \dd x^A \dd x^B \,,  
\label{def:physgnc} 
\een 
on the corresponding region in the physical spacetime $(M, g_{ab})$. 
% \footnote{ %%% 
% For example, when $r= 1/\Omega $, $\tilde \alpha = - \Omega^2$, 
% $\tilde \beta_A =0$, and $\tilde \gamma_{AB} = \sigma_{AB} $(the spherical 
% metric), the above metric describes a conform ally Minkowski metric, 
% \ben
% \dd \tilde s^2 = r^{-2} \{ - \dd u^2 + 2\dd u \dd r  + r^2 \dd \sigma^2\}\,. 
% \een
% } 
This coordinate system in $(M,g_{ab})$ is a sub-class of 
Bondi coordinates~\cite{BBM,Sachs62}.

\medskip 
The asymptotic flatness conditions given in eqs.~(3) and (4) 
are now expressed in terms of the Gaussian null coordinates, 
eq.~(\ref{def:gnc}) above, as 
\bena
&&
\tilde \gamma_{AB} = \sigma_{AB} + O(\Omega^{(d-2)/2}) \,, \quad  
\tilde \gamma^{AB} \frac{\partial}{\partial u} \tilde \gamma_{AB} 
         = O(\Omega^{d/2}) \,,\,\, 
\tilde \gamma^{AB} \frac{\partial}{\partial \Omega}\tilde \gamma_{AB} 
         = O(\Omega^{(d-2)/2}) 
\non  
\\
&&
\tilde \beta_A\,, \,\, \tilde \beta^A  =  O(\Omega^{d/2}) \,, \quad
\tilde \alpha = -\Omega^2 + O(\Omega^{(d+2)/2}) \,.   
\label{condi:flat:gnc}
\eena  
%  
%\footnote{ %%% 
% Note that as already commented below eq.~(\ref{condi:asympt-flat}), 
% the asymptotic flatness (fall-off) conditions given above are derived 
% based on linear perturbation analysis. It might be possible that higher 
% order perturbation effects alter the asymptotic fall-off behavior of, 
% ${\tilde \alpha}, {\tilde \beta}_A, {\tilde \gamma}_{AB}$. 
% In this paper, however, we are not going to pursue higher order 
% effects on the asymptotic flatness conditions. 
%} %%% 
% 

\bigskip 
% \subsection{Asymptotic translational symmetries: no super-translation} 
% An asymptotic symmetry is a diffeomorphism $\phi$ that maps any 
% asymptotically flat metric $g_{ab}$ to an asymptotically flat metric. 

% Although basic properties of asymptotic symmetries in higher dimensions 
% have been studied in \cite{HI05} in the Minkowskian conformal gauge, 
It is worth re-examining asymptotic translational symmetries in terms of 
the Gaussian null coordinate system, as it helps to manifest difference 
between symmetry aspects in $4$-dimensions and in higher dimensions. 
Let $\xi^a$ be a generator of asymptotic symmetries at null infinity, 
and let $\tilde{\chi}_{ab} \equiv %\Omega^{(d-2)/2}\chi_{ab} 
\Omega^2 \pounds_\xi g_{ab}$. Then, for any asymptotically flat metric 
$g_{ab}$, $\tilde{\chi}_{ab}$ must satisfy our 
asymptotic flatness conditions, eq.~(\ref{condi:asympt-flat}). 
Thus, in particular, the relevant components, ${\tilde \chi}_{\mu \nu}$, 
of ${\tilde \chi}_{ab}$ in the Gaussian null conformal gauge must satisfy   
\bena 
 \tilde{\chi}_{u u} = O(\Omega^{(d+2)/2}) \,, \quad 
 \tilde{\chi}_{u A} = O(\Omega^{d/2}) \,, \quad 
 \tilde{\chi}_{AB} = O(\Omega^{(d-2)/2}) \,, \quad 
 {\bar g}^{\mu \nu} \tilde{\chi}_{\mu \nu} = O(\Omega^{d/2}) \,. 
\label{condi:af}
\eena  
Note that for $d=4$, the last condition should be replaced 
with ${\bar g}^{\mu \nu} {\tilde \chi}_{\mu \nu} = O(\Omega)$.

\medskip 
Let us consider a vector field $\xi^a = {\tilde \xi}^a$ in 
$({\tilde M}, {\tilde g}_{ab})$ of the following form~\cite{GW81},  
\ben
 \xi^a = \tau \tnabla^a \Omega - \Omega \tnabla^a \tau \,, 
\een  
where $\tau$ is some function of the {\sl angular} coordinates $x^A$. 
Then, we find 
\bena
&& {\tilde \chi}_{uu} = O(\Omega^{(d+2)/2}) \,, 
\quad 
 {\tilde \chi}_{uA} = O(\Omega^{d/2}) \,, 
\quad 
\non 
\\
&&  {\tilde \chi}_{AB} 
% = 
%  -2 \Omega D_AD_B \tau 
%  + \Omega^2 \frac{\partial }{\partial \Omega}
%            \left(\frac{{\tilde \gamma}_{AB}}{\Omega^2}\right)
%         \left\{
%                ( {\tilde \beta}^C {\tilde \beta}_C - \tilde \alpha )\tau 
%                + \Omega {\tilde \beta}^C D_C \tau 
%         \right\}  
%   + \tau \frac{\partial {\tilde \gamma}_{AB}}{\partial u} 
%   -2\tau D_{(A} {\tilde \beta}_{B)} 
% \non \\ 
 = -2 \Omega \left( D_A D_B + \sigma_{AB} \right) \tau + O(\Omega^{(d-2)/2})
\,, \quad 
 {\tilde g}^{\mu \nu}\tilde{\chi}_{\mu \nu} 
 = {\tilde \gamma}^{AB}{\tilde \chi}_{AB} + O(\Omega^{d/2}) \,,   
\eena 
where $D_A$ denotes the derivative operator with respect to 
${\tilde \gamma}_{AB}$.  

\medskip 
It is clear that when $d=4$, 
% for both ${}^1\!{\bar \xi}$ and ${}^2\!{\bar \xi}$, 
${\tilde \chi}_{\mu \nu}$ satisfy the asymptotic flatness conditions, 
eq.~(\ref{condi:af}) with 
${\bar g}^{\mu \nu} {\tilde \chi}_{\mu \nu} = O(\Omega)$.  
% for an arbitrary function $\tau$.  
Therefore % both ${}^1\!{\bar \xi}$ and ${}^2\!{\bar \xi}$ 
${\tilde \xi}^a$ is an (infinitesimal) asymptotic symmetry.  
% (and they are equivalent as an asymptotic symmetry).
In fact, $\tau$ can be taken as an arbitrary function on $2$-dimensional 
sphere, and thus there exist infinitely many, different 
(angle-dependent) symmetry generators, 
i.e., {\sl supertranslations}~\cite{BBM,Sachs62b}. 
% As is well-known, asymptotic symmetries that correspond to the ordinary 
% four translational symmetries are given by taking $\tau$ 
% as the spherical harmonics on $2$-sphere with the angular quantum numbers 
% $l=0$ and $l=1$ (for which, all the components, $\chi_{\mu \nu}$, 
% vanish identically and ${}^1\!{\bar \xi}$ and ${}^2\!{\bar \xi}$ 
% are exact, translational Killing symmetries).   

\medskip 
In contrast, when $d>4$, 
% both ${}^1\!{\bar \xi}$ and ${}^2\!{\bar \xi}$ 
${\tilde \xi}^a$ fails, in general, to be an infinitesimal asymptotic 
symmetry, due to the first term of ${\tilde \chi}_{AB}$. 
% two cases: 
% (i) When $\tau = const.$, $\tilde{\chi}_{\mu \nu}=0$ and 
% ${}^1\!{\bar \xi}$ describes an exact time-translational symmetry, and   
% (ii) when 
However, if $\tau$ is a spherical harmonic function on $(d-2)$-sphere 
with the second lowest angle quantum number, i.e., 
if $\tau$ is a solution to $\{\sigma^{AB}D_AD_B + (d-2) \}\tau = 0$, 
then it follows that $(D_AD_B+ \sigma_{AB})\tau = O(\Omega^{(d-2)/2})$, 
and therefore that ${\tilde \chi}_{AB} $ and 
${\tilde g}^{\mu \nu} {\tilde \chi}_{\mu \nu}$ 
also satisfy the asymptotic flatness conditions, eq.~(\ref{condi:af}).    
Hence in this case, ${\tilde \xi}^a$ can become an infinitesimal asymptotic 
symmetry. 
% Since there are $d-1$ independent spherical harmonics with the angular 
% quantum number $l=1$ on $d-2$-dimensional sphere, 
% together with case (i) ($l=0$, $\tau = const.$), we have $d$ 
% independent, translational symmetries. 
Since % both ${}^1\!{\bar \xi}$ and 
${\tilde \xi}^a$ cannot be an infinitesimal asymptotic symmetry 
for general $\tau$, there are no supertranslations in higher dimensions,  
as pointed out in \cite{HI05} under the Minkowskian conformal gauge.

\subsection{Bondi energy in Gaussian null conformal gauge}
%---------------------------------------------------------------------------

In the Gaussian null conformal gauge, the background curvature is 
non-vanishing, $ {\bar S}_{ab} \neq 0$, 
% $ {\bar S}_{ab} = - \bar{g}_{ab} +2\sigma_{ab} \neq 0$, 
and instead of eq.~(\ref{condi:flatgauge}), we have   
\bena  
{\tilde S}_{ab} 
   &=& \left( \Omega^2 + O(\Omega^{d/2}) \right) 
            {\tilde \nabla}_{(a}u {\tilde \nabla}_{b)}u 
    + \left( - 2 + O(\Omega^{(d-2)/2}) \right) 
            {\tilde \nabla}_{(a} u {\tilde \nabla}_{b)}\Omega
 \non \\  
   && + O(\Omega^{d/2}){\tilde \nabla}_{(a}u {\tilde \nabla}_{b)}x^A 
      + O(\Omega^{(d-4)/2})  
      \left( 
            {\tilde \nabla}_{(a} \Omega {\tilde \nabla}_{b)}x^A 
            + {\tilde \nabla}_{(a}x^A {\tilde \nabla}_{b)}x^B  
      \right) \,,  
\label{gnc:Sab}
\\ 
{\tilde n}^a {\tilde n}_a &=& \Omega^2 + O(\Omega^{(d+2)/2}) \,, 
\label{gnc:nn}
\\ 
{\tilde \nabla}_a {\tilde n}_b 
% = - {\tilde \Gamma}^\Omega_{ab} 
  &=&  \left( - \Omega^3 + O(\Omega^{(d+2)/2}) \right) 
            {\tilde \nabla}_{a}u {\tilde \nabla}_{b}u  
    + \left( 2 \Omega + O(\Omega^{d/2}) \right) 
            {\tilde \nabla}_{(a}u {\tilde \nabla}_{b)}\Omega 
\non \\ 
 && + O(\Omega^{(d+2)/2}) {\tilde \nabla}_{(a}u {\tilde \nabla}_{b)}x^A 
    + O(\Omega^{(d-2)/2}) 
      \left( {\tilde \nabla}_{(a}\Omega {\tilde \nabla}_{b)}x^A  
             + {\tilde \nabla}_{(a}x^A {\tilde \nabla}_{b)}x^B  
      \right) \,. 
\label{gnc:nabla-n}
\eena
% \bena
% \boxed{
% \begin{aligned}
% &{} \quad \\ 
% &{}\quad {\tilde \Gamma}^\Omega_{uu} &=& \; \Omega^3 + O(\Omega^{(d+2)/2})   
% \quad \non \\
% &{} \quad {\tilde \Gamma}^\Omega_{u\Omega} &=& \; - \Omega + O(\Omega^{d/2}) 
% \quad \non \\
% &{} \quad {\tilde \Gamma}^\Omega_{uA} &=& \; O(\Omega^{(d+2)/2}) 
% \quad \non \\
% &{} \quad {\tilde \Gamma}^\Omega_{\Omega A} \,, \; 
%     {\tilde \Gamma}^\Omega_{AB} &=&  \; O(\Omega^{(d-2)/2}) 
% \quad \non \\
% &{} \quad {\tilde \Gamma}^\Omega_{\Omega \Omega} &=& \; 0 
% \quad \\
% &{} \quad 
% \end{aligned}
% } 
% \eena
It is immediately seen that the news tensor considered above, 
eq.~(\ref{def:news:flat}), becomes singular in $d>4$ on this gauge choice. 
For example, when the physical spacetime is 
Minkowskian, $\tg_{ab}= {\bar g}_{ab}$, and the news, 
eq.~(\ref{def:news:flat}), 
% under the Minkowskian gauge 
becomes $N_{ab}= \Omega^{-(d-4)/2} \sigma_{ab}$, 
which is singular for $d>4$ and also (even in $d=4$) fails to possess 
the desired property that news tensor should vanish for 
any stationary spacetime in the present gauge.

\medskip 
From the observations above, we define a {\sl regularized} news tensor 
under the Gaussian null conformal gauge by subtracting $\sigma_{ab}$ from 
eq.~(\ref{def:news:flat}), so that 
\ben 
 N_{ab} \equiv % \zeta^* \left(\Omega^{-(d-4)/2} \Delta S_{ab}\right) = 
   \zeta^* \left( \Omega^{-(d-4)/2} q^m{}_a q^n{}_b {\tilde S}_{mn} \right) 
          - \Omega^{-(d-4)/2} \sigma_{ab} 
         = \zeta^* \left(\Omega^{-(d-4)/2} \Delta S_{ab}\right)\,,    
\label{def:news:gnc}
\een 
where $\Delta S_{ab} \equiv \tilde S_{ab} - \bar S_{ab}$. 
This is a global definition as it involves the $(d-2)$-dimensional 
sphere metric $\sigma_{ab}$, which specifies a background structure 
at $\I$ and looks a natural higher dimensional generalization of 
the news tensor, $N_{ab}= \zeta^* ( {\tilde S}_{ab}) - \rho_{ab}$, 
in $4$-dimensions, given by Geroch~\cite{Geroch77}.   
(In $4$-dimensional case, the global specification, $\rho_{ab}$,  
is defined by equation~(33) in \cite{Geroch77} and can be taken 
such that $\rho_{ab} = \sigma_{ab}$ \cite{Wald-Zoupas00}.)
%In terms of $\Delta S_{ab} = \tilde S_{ab} - \bar S_{ab}$, $N_{ab}$ is 
%expressed as 
%$N_{ab} = \zeta^* \left(\Omega^{-(d-4)/2} \Delta S_{ab}\right)$. 
It follows from the asymptotic flatness conditions and the vacuum Einstein's 
equations that $\Delta S_{uu} \,, \; \Delta S_{u A} = O(\Omega^{d/2})$ and 
$\Delta S_{AB} = O(\Omega^{(d-4)/2})$, and therefore $N_{AB} = O(1)$. 
Also it can be checked under the Gaussian null conformal gauge that 
when the spacetime considered is stationary, 
(i.e., $\partial /\partial u$ is a Killing field) 
the news, eq.~(\ref{def:news:gnc}), is vanishing, as desired. 
Furthermore, from the $(\Omega, \: \Omega)$-component of 
the vacuum Einstein equations  
% \ben
%  0 = 
%     \frac{\partial}{\partial \Omega} \left({\tilde \gamma}^{AB}
%     \frac{\partial}{\partial \Omega}{\tilde \gamma}_{AB} \right) 
%     + \half {\tilde \gamma}^{CA} {\tilde \gamma}^{BD}
%       \left(\frac{\partial}{\partial \Omega}{\tilde \gamma}_{AB}\right)
%       \left(\frac{\partial}{\partial \Omega}{\tilde \gamma}_{CD}\right) \,, 
%\een  
and the asymptotic flatness conditions above, it can be shown that 
$\Omega^{-(d-4)}q^{bd}q^{ce}\sigma_{de} \tnabla_b \tell_c = O(\Omega)$. 
% $
%  2q^{bd}q^{ce}\sigma_{de} \tnabla_b \tell_c 
%  = \gamma^{AB}{\partial \gamma_{AB} }/{\partial \Omega}
%  = O(\Omega^{d-3}) 
% $. 
Thus, one can anticipate that the first term in the right-hand side of 
eq.~(\ref{def:P:Minkowskiangauge}), 
% $\Omega^{-(d-4)/2}q^{bd}q^{ce}\tS_{de}$ in the Bondi energy-momentum, 
% defined under the Minkowskian conformal gauge 
can simply be rewritten in terms of the regularized news tensor 
defined above, eq.~(\ref{def:news:gnc}). 
%Namely, in a neighborhood of $\I$ 
%we introduce, as candidate for the Bondi energy-momentum, the following 
%vector field, 

\medskip 
From the observations above, we expect the following vector field 
would be a possible candidate for higher dimensional Bondi energy-momentum  
with respect to asymptotic time-translations $\xi^a = \tau \tn^a$ 
(with $\tau = const.$ on $\I$), 
\ben
 P^a[\xi] 
 \equiv \frac{1}{8(d-3)\pi G} 
   \left( 
        \Omega^{-(d-4)/2} 
          \xi^{[a} q^{b]d} N_{de} q^{ce} C^f{}_{bc} \tell_f 
         - \Omega^{-(d-3)} {\tilde C}^{abcd} \xi_b \tell_c n_d 
   \right) \,,   
\label{P}  
\een 
where $C^c{}_{ab} $ is the connection defined by,  
$
({\tilde \nabla}_a - \bar{\nabla}_a)\omega_b = C^c{}_{ab} \omega_c 
$, for an arbitrary $1$-form $\omega_c$. 
Note that in $4$-dimensions, the above formula, eq.~(\ref{P}), 
agrees with the known, $4$-dimensional Bondi energy-momentum integrand.   
% The formula, eq.~(\ref{P}) takes the universal form in arbitrary 
% even spacetime dimensions. 
% 
We now state our main results: % in the following theorem. 
% 

% \bigskip 
\paragraph{Theorem.} 
Consider an even-dimensional $d\geqslant 4$ vacuum spacetime $(M,g_{ab})$  
which is asymptotically flat at null infinity $\I$, satisfying 
the boundary conditions, eq.~(\ref{condi:asympt-flat}) 
(or equivalently eq.~(\ref{condi:af:linear}), and 
eq.~(\ref{condi:af:linear:4d}) for $d=4$). 
Let ${\cal O} \subset {\tilde M} $ be a neighborhood of $\I$ in which 
the Gaussian null conformal gauge, eq.~(\ref{def:gnc}), can be taken. 
In $\cal O$, the divergence, $\tnabla_aP^a$, of $P^a$ introduced 
by eq.~(\ref{P}) smoothly extends to $\I$ and $P^a$ 
solves eq.~(\ref{def:eq:P}). 
% 

% \bigskip 
\paragraph{Proof.} 
%\noindent
%{\bf Proof.} 
We proceed along a similar line of Appendix~A of \cite{HI05}. 
However, this time we need to treat background curvature quantities 
with more care since they are, different from the case of 
the Minkowskian conformal gauge, no longer vanishing upon the Gaussian 
null conformal gauge condition.   
%
% eq.~(\ref{gauge:flat}).)     
% 
% The first few steps parallel the analysis around eqs.~(63), (70), (71), 
% and (82) of \cite{HI05}. 
%**************
%
% We find, instead of eq.~(83) of \cite{HI05}, 
% that the symplectic potential $(d-1)$-form with respect to 
% $\xi^a = \tau {\tilde n}^a$ is written, in terms of the regularized 
% news tensor, $N_{ab}$, eq.~(\ref{def:news:gnc}), as 
% \bena 
%   \Theta(g; \pounds_{\tau n} g) 
%  &=& \frac{1}{32\pi G} \tau \Omega^{-(d-4)} 
%    \zeta^*\left( 
%                 {\Delta S}_{ab} {\Delta S}_{cd} q^{ac}q^{bd} 
%           \right) \cdot \tilde \epsilon 
%\non \\
%   &=& \frac{1}{32\pi G} \tau \Omega^{-(d-4)} N^{ab}N_{ab} \cdot 
%       \tilde \epsilon \,.  
%\label{Theta:NN}
%\eena
%Therefore, equation~(\ref{Theta:NN}), together with the flux formula, 
%eq.~(\ref{def:flux}), yields that the energy flux, $F_\xi$, 
%through $\I$ is always negative. 
%
%**************
%
%\subsection{Bondi energy-momentum in Gaussian null conformal gauge}  
%\medskip 
% Now we wish to derive an expression of Bondi energy-momentum vector, 
% $P^a$, as a solution to eq.~(\ref{def:eq:P}) under the Gaussian 
% null conformal gauge and define the Bondi energy by eq.~(\ref{def:H}).  
In the following, for definiteness we choose 
$\tell_a = (\dd u)_a$, 
so that $\tell^a = (\partial /\partial \Omega)^a \: 
(= \bar \ell^a)$ is a past 
directed (when future null infinity, $\I^+$, is concerned) 
null vector that satisfies 
\ben 
 {\tilde \nabla}_a \tell_b 
 = - \Omega \tell_a \tell_b + O(\Omega^{(d-4)/2}) \,, 
\quad 
 \tell^a \tell_a = 0 \,, 
\quad 
 \tell^a \tn_a = 1 \,,  
\een  
with $\Omega$ being an affine parameter. (Note that 
$ q^{ac}{\tilde \nabla}_b\tell_c = q^{ad}C^c{}_{bd} \tell_c +O(\Omega)$. 
So, when $C^a{}_{bc}$ appears with the contraction with $q^{ab}$ and 
$\tell_a$, we can use $C^a{}_{bc}$ and $\tilde \nabla_a \tell_b$ 
interchangeably.)   
% and as in \cite{HI05}, $q_{ab} = \tilde{g}_{ab} -2\tell_{(a} n_{b)}$.   

\medskip 
Let us consider the identity 
\bena
 2 \tn^a \tnabla_{[a} \tnabla_{b]}\tell_c 
 &=& \tilR_{abcd}\tn^a\tell^d 
%\non \\
% &=& 
  = \tC_{abcd}\tn^a\tell^d 
   + \half {\tilde S}_{bd}\tn_c \tell^d 
   - \half {\tS}_{bc} 
   - \half \tell_b \tnabla_c \tf 
   + \half \tg_{bc}\tell^d\tnabla_d \tf \,,     
\eena
where $\tilde f= \Omega^{-1}\tn^a n_a$. 
Subtracting $2\tn^a \tnabla_{[a} {\bar \nabla}_{b]}\tell_c$ from 
the above equation and using ${\bar C}_{abcd}=0$ and $ 
 {\bar R}_{abcd} = {\bar g}_{ad}\left({\bar g}_{bc}-\sigma_{bc}\right)
                 - {\bar g}_{ca}\left({\bar g}_{bd}-\sigma_{bd}\right)
                 + {\bar g}_{bd}\sigma_{ca} - {\bar g}_{bc}\sigma_{da}
$ for our background geometry, eq.~(\ref{def:bckgrd:gnc}), 
% \ben
%  2\tn^a \tnabla_{[a} {\bar \nabla}_{b]}\tell_c 
%  = \half \bar{S}_{bd}n_c\tell^d - \half \bar{S}_{bc} 
%     - \half \tell_b {\bar \nabla}_c{\bar f} 
%     + \half {\bar g}_{bc} {\bar \tell}^d {\bar \nabla}_d {\bar f} \,.  
% \een 
we have 
\bena
% && (A3) \Leftrightarrow \non \\ 
 2\tn^a \tnabla_{[a} \left( C^e{}_{b]c} \tell_e \right) 
 &=& \tC_{abcd}\tn^a\tell^d 
   + \half \Delta S_{bd}\tn_c \tell^d 
   - \half \Delta S_{bc} 
   - \half \tell_b \tnabla_c \Delta f 
  + O(\Omega^{(d-2)/2}) \,,
\label{eq:A3}
\eena
with $ \Delta f= \Omega^{-1}(\tn^a - {\bar n}^a)n_a$. 
Multiplying $2\Omega^{-(d-4)} \Delta S_{de}q^{bd}q^{ce}$, which is 
$O(\Omega^{-(d-4)/2})$, to eq.~(\ref{eq:A3}), we have 
% \bena  
%  4\Omega^{-(d-4)}\Delta S_{de}q^{bd}q^{ce}\tn^a\tnabla_{[a}C^e{}_{b]c}\tell_e
% = 2 \Omega^{-(d-4)} \tC_{abcf}\tn^a\ell^f \Delta S_{de} q^{bd}q^{ce} 
% - \Omega^{-(d-4)} \Delta S_{ab} \Delta S_{cd} q^{ac}q^{bd} 
% + O(\Omega) \,.
%\eena
% Therefore 
\bena
% && (A4) \Leftrightarrow \non \\ 
  \Omega^{-(d-4)}\Delta S_{ab}\Delta S_{cd}q^{ac}q^{bd} 
  &=& - 4\Omega^{-(d-4)}\Delta S_{de}q^{bd}q^{ce} 
      \tn^a \tnabla_{[a} \left( C^f{}_{b]c}\tell_f \right) 
\non \\ 
&& 
    + 2\Omega^{-(d-4)}\tC_{abcf} \tn^a\tell^f \Delta S_{de} q^{bd}q^{ce} 
    + O(\Omega) \,. 
\label{eq:A4}
\eena 
Using the formulas,  
$
 \tC_{abcf} \tn^a = \Omega \tnabla_{[c}\tS_{f]b} 
                  = \Omega \tnabla_{[c} \Delta S_{f]b} 
                  + \Omega C^e{}_{b[c} {\bar S}_{f]e} 
$, $C^a{}_{bc} = O(\Omega^{(d-4)/2}) = \Delta S_{ab}$,  
and $\tn_cq^{ce} = O(\Omega^2)$, the second term in the right-hand side of 
eq.~(\ref{eq:A4}) above is rewritten as 
\bena
% && (A5) \Leftrightarrow \non \\ 
&& 2\Omega^{-(d-4)}\tC_{abcf} \tn^a\tell^f \Delta S_{de} q^{bd}q^{ce} 
  = - \frac{d-4}{2}\Omega^{-(d-4)}\Delta S_{ab}\Delta S_{cd} q^{ca}q^{bd} 
    + O(\Omega) \,.
\label{eq:A5}
\eena
Next, we rewrite the first term of the right-hand side of eq.~(\ref{eq:A4}) 
using the following formula  
\bena 
&& \tnabla_a \left\{ 
                  \Omega^{-(d-4)} \tn^{[a} q^{b]d}\Delta S_{de} 
                  q^{ce}C^e{}_{bc} \tell_e 
           \right\} 
\non \\
&& \qquad 
 = \Omega^{-(d-4)} \Delta S_{de} q^{ce}q^{bd} \tn^{a} \tnabla_{[a}
                                              \left(C^f{}_{b]c}\tell_f \right)
  + \tnabla_a\left\{
                    \Omega^{-(d-4)}\tn^{[a}q^{b]d}\Delta S_{de}q^{ce}
             \right\} C^e{}_{bc} \tell_e 
\non \\
 && \qquad = \Omega^{-(d-4)} \Delta S_{de} q^{ce} q^{bd} \tn^{a} \tnabla_{[a} 
                                              \left(C^f{}_{b]c}\tell_f \right) 
   + \Omega^{-(d-4)} \tn^{[a}q^{b]d}(\tnabla_a \Delta S_{de})
                     q^{ce} C^f{}_{bc}\tell_f 
   + O(\Omega) \,, 
\eena      
where we have used $\tn_a\tn^a = O(\Omega^2)=\tn_aq^{ab}$,  
$\tnabla_a \tn^a = O(\Omega) = \tnabla_a q^{bc}$.  
Thus, we have  
\bena
%  && (A7) \Leftrightarrow \non \\ 
&& -4 \Omega^{-(d-4)} \Delta S_{de} q^{bd}q^{ce} 
                   \tn^a \tnabla_{[a}\left(C^e{}_{b]c} \tell_e \right) 
     = - 4\tnabla_a \left\{
                          \Omega^{-(d-4)} \tn^{[a}q^{b]d} \Delta S_{de} 
                          q^{ce} C^f{}_{bc}\tell_f 
                   \right\} 
\non \\
 && \qquad \qquad \qquad \qquad \qquad 
      + 4 \Omega^{-(d-4)} \tn^{[a}q^{b]d} 
          \left( \tnabla_a \Delta S_{de} \right) q^{ce}C^f{}_{bc} \tell_f 
      + O(\Omega) \,.  
\label{eq:A7}  
\eena 
Also, from $\Delta S^m{}_m = O(\Omega^{(d-2)/2})$, 
$q^{bd}q^{ce}\tnabla_d\tnabla_e \Delta f = O(\Omega^{d/2})$, 
$\tell^d\tnabla_d \Delta f = O(\Omega^{(d-2)/2})$, 
it follows that 
\bena
  -4 \Omega^{-(d-4)}\tn^{[a}q^{b]d}\left(\tnabla_e \Delta S_{ad} \right)
                 q^{ce}C^f{}_{bc} \tell_f 
  = O(\Omega) \,. 
\eena 
Therefore, adding this to the right-hand side of eq.~(\ref{eq:A7}),  
we can antisymmetrize the second term of eq.~(\ref{eq:A7}) 
with respect to the indices $a$ and $e$ and thereby obtain   
\bena
% && (A7) \Leftrightarrow \non \\ 
&& -4 \Omega^{-(d-4)} \Delta S_{de} q^{bd}q^{ce} 
                   \tn^a \tnabla_{[a}\left(C^e{}_{b]c} \tell_e \right) 
\non \\ && 
     = - 4\tnabla_a \left\{
                          \Omega^{-(d-4)} \tn^{[a}q^{b]d} \Delta S_{de} 
                          q^{ce} C^f{}_{bc}\tell_f 
                   \right\} 
      + 8 \Omega^{-(d-4)} \tn^{[a}q^{b]d} 
          \left( \tnabla_{[a} \Delta S_{e]d} \right) q^{ce}C^f{}_{bc} \tell_f 
      + O(\Omega) \,.  
\eena 
Then, inserting into the above equation 
\ben
  \tnabla_{[a}\Delta S_{e]d} = - \Omega^{-1}\tC_{aedf}\tn^f 
                               - C^f{}_{d[a}{\bar S}_{e]f} \,, 
\een 
we have 
\bena
 -4 \Omega^{-(d-4)} \Delta S_{de} q^{bd}q^{ce} 
                   \tn^a \tnabla_{[a}\left(C^e{}_{b]c} \tell_e \right) 
 &=& - 4\tnabla_a \left\{
                          \Omega^{-(d-4)} \tn^{[a}q^{b]d} \Delta S_{de} 
                          q^{ce} C^f{}_{bc}\tell_f 
                   \right\} 
\non \\ 
 &-& 8 \Omega^{-(d-3)} \tn^{[a}q^{b]d}  
                          \tC_{aedf} \tn^f q^{ce}C^g{}_{bc} \tell_g 
\non \\ 
  &-& 8 \Omega^{-(d-4)} \tn^{[a}q^{b]d} 
          C^f{}_{d[a} {\bar S}_{e]f} q^{ce}C^g{}_{bc} \tell_g 
      + O(\Omega) \,.  
\label{eq:toward:A10}  
\eena 
The third term of the right-hand side of eq.~(\ref{eq:toward:A10}) 
turns out to be $O(\Omega)$ as shown in Appendix.  
The second term of the right-hand side of eq.~(\ref{eq:toward:A10}) is 
(see Appendix, for the derivation) 
\bena
&& - 8 \Omega^{-(d-3)} \tn^{[a}q^{b]d}  
                          \tC_{aedf} \tn^f q^{ce}C^g{}_{bc} \tell_g 
\non \\ 
 && \qquad 
 = 
 - 4\tnabla_{d}\left\{
                     \Omega^{-(d-3)} \tC^{aedf}n_a \tell_e n_f 
               \right\} 
 + 4\Omega^{-(d-3)} \tC_{ae}{}^d{}_f \tn^a\tell^e \tn^f\tn^c\tnabla_d \tell_c 
 + 4\Omega^{-(d-3)} \tC^{aedf} \tell_e n_f \tnabla_d n_a \,.  
\eena
Thus, we find that the first term of the right-hand side of eq.~(\ref{eq:A4}) 
becomes 
\bena
 - 4\Omega^{-(d-4)}\Delta S_{de}q^{bd}q^{ce} 
      \tn^a \tnabla_{[a} \left( C^f{}_{b]c}\tell_f \right) 
 &=& - 4\tnabla_{d}
     \left\{
        \Omega^{-(d-4)} \tn^{[a}q^{b]d}\Delta S_{de} q^{ce} C^f{}_{bc} \tell_f 
        -\Omega^{-(d-3)} \tC^{abcd}n_b \tell_c n_d 
     \right\} 
\non \\
 &+& 4\Omega^{-(d-3)} \tC^{aedf} 
   \left(
           \tn_a \tell_e \tn_f \tn^c\tnabla_d \tell_c 
         + \tell_e n_f \tnabla_d n_a 
   \right) 
\non \\
 &-& 8 \Omega^{-(d-4)}\tn^{[a}q^{b]d} C^f{}_{d[a} {\bar S}_{e]f} 
                     q^{ce}C^g{}_{bc} \tell_g  
  + O(\Omega) \,. 
\label{term:1st:rhsA4}  
\eena 
Substituting eqs.~(\ref{eq:A5}) and (\ref{term:1st:rhsA4}) 
into eq.~(\ref{eq:A4}), we have 
\bena
 (d-3)\Omega^{-(d-4)} \Delta S_{ab} \Delta S_{cd}q^{ca}q^{bd} 
 &=& - 4\tnabla_a 
     \left\{ 
        \Omega^{-(d-4)} \tn^{[a}q^{b]d}\Delta S_{de} q^{ce} C^f{}_{bc} \tell_f 
        -\Omega^{-(d-3)} \tC^{abcd}n_b \tell_c n_d 
     \right\}  
\non \\
 && + 2\Omega^{-(d-4)} \tC_{abcf}\tn^a\tell^f \tg^{bd} \tg^{ce}{\bar S}_{de} 
    + 4\Omega^{-(d-3)} \tC_{aedf}\tn^a\tell^e \tn^f\tn^c \tnabla^d \tell_c 
\non \\
 && - 8 \Omega^{-(d-4)}\tn^{[a}q^{b]d} C^f{}_{d[a} {\bar S}_{e]f} 
                     q^{ce}C^g{}_{bc} \tell_g  \,. 
\label{eq:50B}
\eena 
The last three terms are shown to be $O(\Omega)$ in Appendix. 
Thus, substituting the defining equation~(\ref{P}) of $P^a$ 
for $\xi^a= \tau \tn^a$, we obtain 
\ben
 \tau \Omega^{-(d-4)} \Delta S_{ab} \Delta S_{cd}q^{ca}q^{bd} 
 = - 32\pi G \tnabla_a P^a + O(\Omega) \,. 
\label{eq:NNP}
\een  
Since %, as discussed below eq.~(\ref{def:news:gnc}), 
$\Delta S_{uu} \,, \; \Delta S_{u A} = O(\Omega^{d/2})$ and 
$\Delta S_{AB} = O(\Omega^{(d-4)/2})$ under our boundary conditions, 
the left-hand side has a smooth extension to $\I$. Hence 
the divergence, $\tnabla_a P^a$, also must smoothly extend to $\I$.  

\medskip 

We also find %, instead of eq.~(83) of \cite{HI05}, 
that the symplectic potential $(d-1)$-form with respect to 
$\xi^a = \tau {\tilde n}^a$ is written, in terms of the regularized 
news tensor, $N_{ab}$, as 
\bena 
  \Theta(g; \pounds_{\tau n} g) 
  &=& \frac{1}{32\pi G} \tau \Omega^{-(d-4)} 
    \zeta^*\left( 
                 {\Delta S}_{ab} {\Delta S}_{cd} q^{ac}q^{bd} 
           \right) \cdot \tilde \epsilon 
%\non \\
%   &=& 
   = \frac{1}{32\pi G} \tau N^{ab}N_{ab} \cdot 
       \tilde \epsilon \,.  
\label{Theta:NN}
\eena
Thus, when $\tau = const.$, comparing eq.~(\ref{eq:NNP}) with 
eq.~(\ref{Theta:NN}) 
% and expressing $\Delta S_{ab}q^{bc}$ in terms of the news tensor, $N_{ab}$, 
we consequently obtain the formula, eq.~(\ref{P}), 
for the translation symmetry $\xi^a= \tau \tn^a$ with $\tau = const$,  
as a solution to eq.~(\ref{def:eq:P}).  
$\Box$ \hfill 

\bigskip 

The theorem above yields that the vector field, $P^a$, defined by 
eq.~(\ref{P}) is a good candidate for the higher dimensional Bondi 
energy-momentum integrand. 
It should be, however, noted that although the integrand of the flux formula, 
eq.~(\ref{def:eq:P}), is well-defined, the limit to $\I$ of $P^a$ itself 
does not appear to exist under our boundary conditions. 
This can be seen by examining the relevant component, $P^u= P^a \tell_a$,  
of eq.~(\ref{P}), 
% 
%\medskip  
% We shall briefly discuss regularity of Bondi energy-momentum vector, 
% analysing gravitational perturbations off the flat background geometry. 
%Although the flux formula---which is given by a divergence of 
%the Bondi energy-momentum vector, $P^a$, (see eq.~(\ref{def:eq:P}))---is 
%well-defined, it is not manifest whether $P^a$ itself, 
%as well as ${\cal H}_\xi$ can be regular at $\I$. 
% Although ${\tilde \nabla}_a P^a$ is regular by definition, $P^a$ itself 
% is in general not manifestly regular in higher dimensions $d>4$. 
%%% 
% The relevant component, $P^u= P^a \tell_a$, of eq.~(\ref{P}) is explicitly 
% written 
% are $P^\Omega = O(\Omega^2)$ and 
\bena 
 P^u[\xi] = \frac{- \tau}{16(d-3)\pi G}
   \Omega^{-(d-3)} 
   \left\{ 
          \frac{1}{2}{\sigma}^{CA}{\sigma}^{DB} 
                     \frac{\partial {\tilde \gamma}_{AB}}{\partial u}
                     \frac{\partial {\tilde \gamma}_{CD}}{\partial \Omega}
         - \frac{\partial}{\partial \Omega} 
           \left[
                 \Omega^2 \frac{\partial}{\partial \Omega} 
                 \left(\frac{{\tilde \alpha}}{\Omega^2}\right) 
           \right]          
   \right\} + O(\Omega) \,.  
\label{comp:Pu} 
%
%\\
% P^A  
% &=& \frac{- \tau}{16(d-3)\pi G}
%   \Omega^{-(d-3)} 
%   \left\{ 
%         \frac{\partial^2 }{\partial u \partial \Omega}{\tilde \beta}^A 
%        + \Omega D^A \frac{\partial }{\partial \Omega}
%                \left( \frac{{\tilde \alpha}}{\Omega} \right) 
%        - {\tilde \alpha} \frac{\partial^2}{\partial \Omega^2} 
%                          {\tilde \beta}^A 
%   \right\} + O(\Omega) \,, 
\eena
% Let us first look at equation for $P^u = P^a \tell_a$.  
%
%\bigskip 
% We shall show below that $P^a$ defined in terms of the regularized 
% news tensor just above is, in fact, a solution to eq.~(\ref{def:eq:P}).  
%%% 
The first term of the right-hand side of above 
eq.~(\ref{comp:Pu})---which comes from the news tensor---is $O(1)$ 
and therefore well-defined, 
% for the Schwarzschild case, 
% as $\tilde \alpha= -\Omega^2 + O(\Omega^{d-1})$ in this case.  
whereas the second term of $P^a\tell_a$ behaves $O(\Omega^{-(d-4)/2})$ 
near $\I$, according to the boundary condition eq.~(\ref{condi:flat:gnc}). 
($P^A$ behaves in a similar way.) Therefore, except the $d=4$ case, 
$P^a$ itself would not appear to be regular at $\I$ for general, 
asymptotically flat radiative spacetimes. 

\medskip 
However, this apparent singular behavior of $P^a$ does not necessarily imply 
that the {\sl integral} of $P^a\tell_a$ over a compact cross-section of $\I$ 
also would be singular for general, asymptotically flat radiative spacetimes.  
% since ${\cal H}_\xi$ is defined as the integral over a compact surface $B$. 
%
Let $(M,g_{ab})$ be a vacuum spacetime which is asymptotically flat 
at future null infinity $\I^+$ and $\Sigma_0$ be a (partial) Cauchy surface 
which extends to {\sl spatial infinity},~$i^0$. 
%$\approx S^{d-2}$ on $\I$. 
Assume further that there is a compact subset $K_0$ of $\Sigma_0$ in $M$ 
such that the region outside the causal future of $K_0$ is stationary, 
and gravitational radiation may present only in $J^+(K_0)$.  
Consider the flux formula, eq.~(\ref{def:flux}), which may be rewritten as 
\bena
 \int_{B} {}^{(d-2)}\!{\tilde \epsilon} \:P^a \tell_a 
 = \int_{B_0} {}^{(d-2)}\!{\tilde \epsilon} \:P^a \tell_a 
  + F_\xi \,,             
%       = - \int_{I_{12}} {\Theta}(g;\pounds_\xi g) 
\label{rel:Bondi:Flux:ADM}
\eena 
where the integral over a given cross-section $B$ of $\I$ 
in eq.~(\ref{rel:Bondi:Flux:ADM}) above is defined by the limitting 
procedure, as described below eq.~(\ref{def:Noethercharge}); we first perform 
the integration over a sequence of $(d-2)$-closed surfaces, $S_i$, 
in some neighborhood of $B$ and then take the limit $S_i \rightarrow B$ 
in a certain manner. 
Since the Gaussian null coordinate chart, $({\cal O},x^\mu)$, is 
now available, one may naturally set $S_\Omega = B(u,\Omega \neq 0)$ 
in $\cal O$ and take the limit $S_\Omega \rightarrow B$ by 
$\Omega \rightarrow 0$ on a $u=const.$ hypersurface. 
(Of course, the integrals in eq.~(\ref{rel:Bondi:Flux:ADM})---as defined on 
$B$ in $\I$, not inside the spacetime $(M,g_{ab})$---should not depend on 
the way how to take $S_i$ in performing the limiting procedure.)
Now let us take $B$ in $\I^+ \cap J^+(K_0,{\tilde M})$ and $B_0$ 
in a sufficiently past so that 
% $B_1 \subset \I^+ \cap \{ {\tilde M} \setminus J^+(K_0,{\tilde M})\} $. 
$B_0 \subset \{ \I^+ \cap D^+ ( \Sigma_0 \setminus K_0,{\tilde M}) \} $. 
Since the spacetime region $D^+(\Sigma_0 \setminus K_0,\tilde M)$ 
is stationary, the integral of $P^a\tell_a$ over $B_0$ is independent of 
the {\sl Killing time} and should correspond to the ADM energy. 
\footnote{ %%%  
For stationary spacetimes with $(\partial/\partial u)^a$ being a timelike 
Killing vector field in a vicinity $\cal O$ of null infinity, 
the news tensor vanishes identically and the integral of $P^a\tell_a$ 
over $B_0$ becomes the integral over $S^{d-2}$ of the {\sl Coulomb part} 
of the Weyl tensor, 
$\propto \Omega^{-(d-3)}{\tilde C}^{abcd}\tell_a \tn_b \tell_c \tn_d$. 
If one can, furthermore, set $\tilde \beta_A=0$ in $\cal O$, then 
it immediately follows from the vacuum Einstein equations, 
the $(u,A)$-components of eq.~(\ref{eq:conf-Einstein}), 
that 
\ben
D_A\Omega^{d-2} \frac{\partial}{\partial \Omega} 
 \left( 
       \frac{{\tilde \alpha}+\Omega^2}{\Omega^{d-2}}   
 \right) = O(\Omega^{d-2}) \,,  
\non 
\een 
and hence that $\tilde \alpha = -\Omega^2+O(\Omega^{d-1})$.   
Thus, in this case, the second term of the right-hand side of 
eq.~(\ref{comp:Pu})---which comes from the Coulomb part 
% $\Omega^{-(d-3)}{\tilde C}^{abcd}\tell_a \tn_b \tell_c \tn_d $ 
of the Weyl tensor---gives rise to a finite, constant value, which is 
to be identified with the ADM energy. We believe that even if 
${\tilde \beta}_A$ is not set to be zero in ${\cal O}$, 
% one would still be able to show that 
the Coulomb part of the Weyl tensor would give rise to the ADM energy 
whenever a cross-section $B_0$ is taken in a portion of $\I^+$ whose 
immediate vicinity $\cal O$ is stationary, 
but we have not fully investigated this case. 
} %%%   
This can be directly verified when the initial data on $\Sigma_0$ 
outside the compact region $K_0$ coincides with Schwarzschild data 
(so that $D^+(\Sigma_0 \setminus K_0,\tilde M)$ is isometric to 
Schwarzschild spacetime).  
Since $\Theta$ is by definition smooth and finite on $\I$, it follows 
from eq.~(\ref{Theta:NN}), together with the flux formula, 
eq.~(\ref{def:flux}), that the energy flux, $F_\xi$, through the segment of 
$\I^+$ with boundaries $B$ and $B_0$ 
% (or equivalently the segment bounded by $B$ and $\partial \Sigma$) 
must be finite. 
Therefore, if the ADM energy evaluated on $B_0$ is finite, then the above 
formula, eq.~(\ref{rel:Bondi:Flux:ADM}), yields that the {\sl integral} of 
$P^a\tell_a$ over the cross-section $B$ also must be finite. 
Furthermore, since the flux $F_\xi$ through a segment of $\I^+$ always is 
negative, as immediately seen from eqs.~(\ref{def:flux}) and~(\ref{Theta:NN}), 
the right-hand side of eq.~(\ref{rel:Bondi:Flux:ADM}) may be viewed as 
the ADM energy minus radiation energy carried away from the spacetime 
by the flux through $\I^+$. Therefore, at least under the setup described 
above, $P^a\tell_a$ given by eq.~(\ref{P}) is indeed our desired 
Bondi energy-momentum integrand.  
More generally, we would now like to make the following conjecture: 

\bigskip \noindent
{\sl %%% 
For any asymptotically flat spacetime (at null infinity $\I$) and 
any compact cross-section $B$ of $\I$, the integral of $P^a\tell_a$ given by 
eq.~(\ref{P}) over a closed surface $S$ always has a well-defined limit as 
$S$ approaches $B$, and the limit is independent of how $S$ approaches $B$.  
} %%%

\bigskip \noindent
Accordingly, we propose that 
{\sl 
${\cal H}_\xi$ defined by eq.~(\ref{def:H}) with the vector 
$P^a$ given by eq.~(\ref{P}) can be taken as the definition for a higher 
dimensional generalization of the Bondi-energy}. 

\bigskip 
% As we have already seen above, it follows from eqs.~(\ref{def:flux}) 
% and (\ref{Theta:NN}) that the energy flux is always negative. 
% Then, equation~(\ref{def:eq:P}), together with 
% the energy-momentum, eq.~(\ref{P}), 
It also follows from eq.~(\ref{rel:Bondi:Flux:ADM}) that 
the Bondi-energy defined by eq.~(\ref{def:H}) is a decreasing 
function in time.  
% 
% (This, of course, also should hold for the formula given 
%  under Minkowskian conformal gauge. But I have not yet seen this.)  
In Gaussian null coordinates, one finds that 
$\tnabla_a P^a = \partial P^u /\partial u + D_A P^A + O(\Omega)$.  
Then, via eqs.~(\ref{def:eq:P}) and (\ref{def:H}), 
the energy loss rate at time $u$ (with the corresponding 
cross-section $B$) is given by 
% the magnitude of {\sl shear}-squared of a congruence of null geodesic 
% generators of $\I$ on $(\tilde M, \tg)$, 
\ben 
% \frac{\tau}{8\pi G} \int_{B} {}^{(d-2)}\!\tilde \epsilon 
% \Omega^{-(d-2)}\sigma^{CA}\sigma^{BD}
%                                    C^\Omega{}_{AB}C^\Omega{}_{CD} 
% =
 \frac{\partial {\cal H}_\xi }{\partial u}
 =  
 - \frac{\tau}{32\pi G}\int_{B} \! {}^{(d-2)}\!\tilde \epsilon \: 
   \Omega^{-(d-2)}\sigma^{CA}\sigma^{DB}
                                    \frac{\partial {\tilde \gamma}_{AB}
                                         }{\partial u} 
                                    \frac{\partial {\tilde \gamma}_{CD}
                                         }{\partial u} \,,     
\label{eq:energyloss}
\een
which corresponds to, e.g., eq.~(5.12) in \cite{{Sachs62}} obtained 
in $4$-dimensions. 

\section{Summary and discussions} 
\label{sect:discussion} 
%---------------------------------------------------------------------------
 
In this paper, we have considered asymptotic flatness at null infinity 
in higher (even spacetime) dimensions in terms of the Gaussian null 
conformal gauge, focusing on vacuum solutions of Einstein's equations.  
As shown in the previous paper \cite{HI05}, when asymptotically flat 
spacetimes are even spacetime dimensional, one can define a smooth 
conformal null infinity $\I$ in the unphysical spacetime 
$(\tilde M, \tg_{ab})$ that is stable against, at least, linear 
perturbations. 
Then, we have discussed that a Gaussian null coordinate system 
$(u,\Omega,x^A)$ with respect to $\I$ can naturally be constructed 
on some neighborhood of $\I$ in $(\tilde M, \tg_{ab})$. 
%
% (A formal proof for the global existence of such a Gaussian null coordinate 
% system needs to be given.)
%
% Therefore it is natural to assume that one can construct 
% a Gaussian null coordinate system with respect to null infinity $\I$ 
% on some neighborhood of $\I$ in the unphysical spacetime 
% $(\tilde M, \tg_{ab})$. 
By taking a luminosity distance as the inverse of 
$\Omega$ (viewed as a smooth, appropriately scaled conformal factor),  
one can define Gaussian null coordinates in the physical spacetime, 
which correspond to a restricted class of Bondi coordinates. 
In contrast to the Minkowskian conformal gauge condition  
employed in \cite{HI05}, the Gaussian null conformal gauge allows us 
to take  compact, spherical cross-sections at $\I$ with $(d-2)$-dimensional 
round sphere metric $\sigma_{ab}$ as a global specification of 
the background structure. Thus, one can directly compare 
the higher dimensional formulas with the $4$-dimensional ones obtained 
in the Bondi coordinates. 
% 
% We have also examined asymptotic translational symmetries under 
The Gaussian null conformal gauge also helps to manifest the difference 
between asymptotic symmetry properties in $4$-dimensions and those 
in higher dimensions, showing clearly the absence of supertranslations 
in higher dimensions. 
% and also shown that $d$ independent spherical harmonics 
% on $(d-2)$-sphere with $d$ lowest angular quantum numbers (i.e., 
% $l=0$ and $l=1$) correspond to the translational Killing symmetries in 
% $d$-dimensional Minkowski spacetime. 

% 
\medskip 
We have modified the definition of the news tensor in higher (even 
spacetime) dimensions so that it becomes regular under the choice 
of the Gaussian null conformal gauge. The new definition of the news 
tensor involves a global specification of the background structure 
$\sigma_{ab}$ on null infinity $\I$, which is not included in the previous 
definition, eq.~(61) of \cite{HI05}.  
% due to the choice of special gauge condition. 
The news tensor defined above, eq.~(\ref{def:news:gnc}), looks natural 
as a higher dimensional generalization of the news tensor 
in $4$-dimensions defined in \cite{Geroch77}. 
Then, for the case of vacuum spacetimes, within the Hamiltonian framework 
of Wald and Zoupas \cite{Wald-Zoupas00}, we have obtained 
the expression of the higher (even) dimensional generalization of 
the Bondi energy, ${\cal H}_\xi$, and 
energy-momentum, $P^a$, for the special asymptotic translation 
$\xi^a = \tau {\tilde n}^a$ with $\tau = const.$ 
in terms of the regularized news tensor, eq.~(\ref{def:news:gnc}). 
% regularized on the Gaussian null conformal gauge. 
\footnote{ %%% 
%
% \medskip 
% The formula, eq.~(\ref{P}), is for the special translation $\xi= \tau \tn^a$ 
% with $\tau = const$. 
The expression for general translational symmetries, 
$\xi^a= \tau \tn^a - \Omega \tnabla^a \tau$, may be obtained by 
noting that if $\phi$ is an asymptotic symmetry, then $\phi_* \xi^a$ 
is also an infinitesimal generator of asymptotic symmetry, 
and considering that the Bondi energy for $\xi^a$ with the metric $g_{ab}$, 
evaluated at $B$ is equal to the Bondi energy for $\phi_* \xi^a$ 
for the metric $\phi^* g_{ab}$, evaluated at $\phi(B)$, as discussed 
in \cite{HI05}. 
} %%% 
% 

% \medskip 
% We have also examined asymptotic translational symmetry under the Gaussian 
% null conformal gauge and have explicitly seen that super-translations do 
% not exist in $d>4$ and also shown that $d$ independent spherical harmonics 
% on $(d-2)$-sphere with $d$ lowest angular quantum numbers (i.e., 
% $l=0$ and $l=1$) correspond to the translational Killing symmetries in 
% $d$-dimensional Minkowski spacetime. 

% 
\medskip  
% We shall briefly discuss regularity of Bondi energy-momentum vector, 
% analysing gravitational perturbations off the flat background geometry. 
We have then pointed out the puzzling fact that although the flux 
formula---which is given in terms of the divergence of $P^a$, 
(see eq.~(\ref{def:eq:P}))---is well-defined, $P^a$ itself would 
not appear to be regular at $\I$ for $d>4$, 
% since it is immediately seen that 
as the second-term of $P^a\tell_a$, eq.~(\ref{comp:Pu}), (and $P^A$, too) 
behaves near $\I$ as $O(\Omega^{-(d-4)/2})$ under our asymptotic 
flatness conditions.   
% and therefore, except the $d=4$ case, 
% $P^a$ itself would appear to be singular at $\I$ for general, 
% asymptotically flat radiative spacetimes. 
%
% \medskip 
We have given an attempt to show that even if $P^a$ itself is 
singular, the integral of $P^a\tell_a$ over a compact cross-section 
would be regular, by discussing a relationship between 
the flux formula, eq.~(\ref{def:flux}) (or eq.~(\ref{rel:Bondi:Flux:ADM})) 
and the ADM energy in the case in which an asymptotically flat, 
radiative spacetime has a stationary region in the past, at least in some 
neighborhood of spatial infinity $i^0$.  
It should be recalled here that the ADM energy is defined at 
{\sl spatial infinity}, $i^0$, with respect to an asymptotic time-translation 
symmetry defined at $i^0$, whereas the Bondi energy is defined at 
a cross-section of {\sl null infinity}, $\I$, with respect to an asymptotic 
time-translation symmetry defined at $\I$. Therefore, in order to find 
a precise relationship between the two notions of gravitational total 
energies, one needs to treat asymptotic properties of gravitational fields 
and asymptotic symmetries at both $i^0$ and $\I$ in a unified manner. 
In the $4$-dimensional case, a framework for such a unified treatment 
of two infinities has been given by \cite{AH78,AMA79}, and 
the interpretation that the Bondi-energy is the ADM energy minus the energy 
carried away by flux through null infinity has been justified~\cite{AMA79}. 
However, the conformal framework of \cite{AH78,AMA79} used to define 
asymptotic flatness at $i^0$ and $\I^+$ in a unified manner is different 
from the present framework for defining asymptotic flatness at null infinity 
under the Gaussian null conformal gauge. For this reason, although we believe 
that the unified treatment of $i^0$ and $\I^+$ given by \cite{AH78,AMA79} 
would be generalized to the higher dimensional case, it would not appear 
to be a straightforward task. 

% However, this does not necessarily imply that the Bondi-energy defined 
% above also would be singular in general radiative spacetimes.   
% since ${\cal H}_\xi$ is defined as the integral over a compact surface $B$. 

\medskip 
Another attempt to justify $P^a$ given by eq.~(\ref{P}) as 
the `legitimate' Bondi energy-momentum integrand may be given by 
considering gravitational perturbations off of Minkowski spacetime. 
The part of $P^a$ that involves the news tensor is $O(1)$, and  
the relevant part to be examined is the Weyl curvature part 
of $P^a \tell_a$ (the second term of the above eq.~(\ref{P})). 
% Let $\Psi = \tC^{abcd}\tn_a \tell_b \tn_c \tell_d $. 
Since the perturbation 
$\Omega^{-(d-3)} \delta \tC^{abcd}\tn_a \tell_b \tn_c \tell_d $ is a scalar 
quantity, it can be expanded in terms of spherical harmonics 
% , ${\Bbb S}_{\bf k}$, defined 
on a $(d-2)$-sphere $B(u,\Omega)$.  
% with ${\bf k}\cdot {\bf k} = l(l+d-3)$ being the eigen value 
% of the harmonics. 
Then, given that the time dependence of perturbations 
% $\tC^{abcd}\tn_a \tell_b \tn_c \tell_d$ 
is $\propto \e^{-i \omega t}$, it turns out from the linearized Einstein 
equations and the Bianchi identities that general solutions 
of perturbations % $\tC^{abcd}\tn_a \tell_b \tn_c \tell_d $ 
are given in terms of Bessel functions and that 
$ \Omega^{-(d-3)} \delta \tC^{abcd}\tn_a \tell_b \tn_c \tell_d $ 
behaves near $\I$ as 
$\sim \Omega^{-(d-4)/2}$, which looks singular. 
% , which would make ${\cal H}_\xi$ singular. 
However, 
% in our Bondi-energy expression, $\tC^{abcd}\tn_a \tell_b \tn_c \tell_d $ 
% will be integrated over a compact cross-section $(d-2)$-sphere, and 
contributions to 
$\Omega^{-(d-3)} \delta \tC^{abcd}\tn_a \tell_b \tn_c \tell_d $ from 
any modes of perturbations, except spherically symmetric mode (the S-wave) 
% with ${\bf k}\cdot {\bf k} \neq 0$ 
vanish identically, when integrated over a compact cross-section 
$(d-2)$-sphere, $B(u)$. 
Therefore only the S-wave becomes relevant. 
It turns out from the Bianchi identity that provided 
$\delta \tC^{abcd}\tn_a \tell_b \tn_c \tell_d$ is not divergent at $\I^+$, 
$\Omega^{-(d-3)} \delta \tC^{abcd}\tn_a \tell_b \tn_c \tell_d $ for the S-wave 
behaves $\sim O(1)$ \cite{Ishibashi-Kodama03}, 
as expected from the uniqueness of static vacuum black holes 
in higher dimensions~\cite{Gibbons}. 
(Of course, the $S$-wave solutions in exact, flat spacetime are singular 
at the centre of the spherical symmetry, but here we are not concerned 
with global regularity of perturbation solutions.) 
Therefore the linear perturbations produce only regular contributions to 
% the Bondi-energy, 
${\cal H}_\xi$. 
% 
%\medskip 
% 
However, this does not seem to be the case for second order perturbations. 
General solutions of second order metric perturbations 
% , $\delta^{(2)}\!g$, 
of a physical spacetime may be given 
% schematically as $ \delta^{(2)}\!g = \int {\cal L}^{-1}S$ 
in terms of the Green's functions, which is the `inverse' of the wave 
operator appearing in the equation of motion for first order 
metric perturbations, together with a source term, given as quadratics 
of the first order perturbations (and their derivatives).  
It turns out from the linearized vacuum Einstein equations that 
% since ${}^{(1)}\!h \sim \Omega^{(d-2)/2}$, together 
% with the volume element in the integration, 
the relevant Green's function consists of powers in $\Omega$, 
spanning all values between $\Omega^{(d-2)/2}$ and $\Omega^{d-3}$ 
(see e.g.,~\cite{CDL03}). 
So, if second order contributions from the second term 
of the right-hand side of eq.~(\ref{comp:Pu}) 
% $\tilde \alpha/\Omega^2$ in the expression of $P^a\tell_a$ 
are determined in this way, then we would have a singular behavior 
in ${\cal H}_\xi$. % (That is quite disappointing, though.) 

\medskip 

%It would not appear to be straightforward to directly check the regularity 
%of $P^a$ and the Bondi energy expression in full non-linear theory.  
In order to fully justify our definition of $P^a$ and the Bondi 
energy expression, more work on non-linear analysis of asymptotic flatness 
conditions needs to be done. For example, one may start with expanding 
the unphysical metric in powers of $\Omega$ at $\I$, and thereby expressing 
the Einstein equations in terms of the expansion coefficients in a similar 
manner performed for spacetimes with a different asymptotic structure 
(see \cite{HIM05} and also references therein), and then we read off 
at which order of $\Omega$ true dynamical information on gravitational 
radiation enters each component of the unphysical metric~\cite{HI-future}.    

% ${\tilde \alpha}, {\tilde \beta}_A, {\tilde \gamma}_{AB}$. 
% In this paper, however, we are not going to pursue higher order 
% effects on the asymptotic flatness conditions. 

\medskip 
Concerning this regularity issue of the Bondi energy-momentum integrand, 
we also should keep in mind that the vector $P^a$ has degrees of 
freedom in the addition of a vector field of the form $\tnabla_b X^{ab}$ 
with $X^{ab}$ being an arbitrary anti-symmetric tensor field on 
$\tilde M$, as commented below eq.~(\ref{def:eq:P}); 
$P'^a = P^a + \tnabla_b X^{ab}$ also solves eq.~(\ref{def:eq:P}) 
if $P^a$ does. A relevant question is then whether there exists a $P'^a$ 
for which $P'^a \tell_a$ smoothly extends to $\I$ and yields an equivalent 
formula for the higher dimensional Bondi-energy. 
We notice that the main cause for the apparent singularity of $P^a\ell_a$ 
is the second term of the right-hand side of eq.~(\ref{comp:Pu}).   
For example, it is not so hard to find such $X^{ab}$ that removes only 
the second term from $P^a\ell_a$ of eq.~(\ref{comp:Pu}), but then 
the Bondi energy expression 
with the new $P'^a= P^a + \tnabla_b X^{ab}$---which is now manifestly 
well-defined at $\I$---would fail to reproduce the ADM energy 
for stationary spacetimes. This problem remains unsolved. 

% In order to fully verify the well-definedness of $P^a$ at $\I$, 
% it is needed to do non-linear analysis. 
%  
%\medskip 
%Together with this regularity problem, the question of whether 
%the higher dimensional Bondi-energy is positive definite is also 
%an important issue left open for future research. 

\medskip 
In some theories of higher dimensional gravity, our $4$-dimensional universe 
is modeled as an embedded (or a boundary) hypersurface---so called 
the {\sl braneworld}---in a higher-dimensional bulk spacetime, 
in which only gravitational radiation can probe extra dimensions.  
Therefore, in this context, it is of considerable interest to define 
the notion of a higher dimensional Bondi-type energy that can be used to 
measure energy flux of gravitational radiation from extra-dimensions. 
% in higher dimensional bulk spacetime, as well as in the $4$-dimensional 
% braneworld. 
For the case in which the bulk spacetime is higher, even-dimensional 
and its curvature radius is sufficiently large compared to typical scale 
of a system of interest (e.g., a mini black hole on the braneworld), 
one may be able to define, within the conformal framework, 
the notion of asymptotic flatness at null infinity $\I$ for both the 
bulk spacetime and the braneworld, simultaneously. In that case, 
our Bondi-energy formula would apply to such a system almost as it stands.  
For example, the flux formula on the $4$-dimensional braneworld 
may simply be given by estimating the boundary integral of the formula, 
eq.~(\ref{rel:Bondi:Flux:ADM}), on a $2$-dimensional closed sub-manifold 
which corresponds to the intersection of $B \subset \I$ and 
the brane's world-volume. 
However, when the bulk spacetime is $5$-(or higher, odd-)dimensional, 
our formula based on the conformal method would not work, as the unphysical 
metric becomes, in general, singular at null infinity. 
For odd-dimensional spacetime, however, it may still be possible to 
define the notion of Bondi-type energy by formulating asymptotic 
structure in terms of the {\sl physical} metric, instead of the conformal 
unphysical metric.  
Also for Kaluza-Klein type models, 
% in which the whole spacetime looks like 
% a product of $4$-dimensional macroscopic spacetime, ${}^{(4)}\!{\cal M}$, 
% and some compactified extra-dimensions, ${\cal K}$. 
the conformal framework would not appear to work. 
One may expect that the null infinity $\I$ of a Kaluza-Klein 
spacetime should be $\I \approx {\Bbb R} \times S^{2} \times {\cal K}$,  
where ${\Bbb R} \times S^{2}$ corresponds to the conformal boundary of 
$4$-dimensional macroscopic spacetime and ${\cal K}$ to the compactified 
extra-dimensions. However, a conformal factor will make ${\cal K}$ 
shrink to a point on $\I$. Again one may be able to define energy flux or 
the Bondi-type energy in Kaluza-Klein spacetime, using the physical metric, 
rather than the unphysical metric. The study of such cases 
with non-trivial asymptotic structure is left open for future work.   

\medskip 
\begin{center}
{\bf Acknowledgements} 
\end{center}

I am greatly indebted to Bob Wald for many useful discussions 
and valuable comments. 
% and guidance along the way. 
I have also benefited greatly from discussions with Stefan Hollands. 
I would also like to thank Chiang-Mei Chen and Jim Nester for discussions. 
I wish to thank Department of Physics, National Central University 
for its hospitality during the time some of this research was carried out. 
This work was supported in part by NSF grant no PHY 04-56619 to 
the University of Chicago.  

\section*{Appendix}  
%---------------------------------------------------------------------------

We first show that the third term of the right-hand side of 
eq.~(\ref{eq:toward:A10}) are $O(\Omega)$. Note first that 
\ben
  C^u{}_{AB} \:,\quad C^A{}_{\Omega B} \: = O(\Omega^{(d-4)/2}) \,,
\een 
and all the other components of the connection $C^a{}_{bc}$ decay faster 
than the above two. Also note that $q^{ab} = 
\sigma^{AB}(\partial/\partial x^A)^a (\partial/\partial x^B)^b 
+ O(\Omega^{(d-2)/2})$. From these, it immediately follows that 
\bena
 && - 8 \Omega^{-(d-4)} \tn^{[a}q^{b]d}C^f{}_{d[a} {\bar S}_{e]f}
                      q^{ce} C^g{}_{bc} \tell_g 
\non \\
&& \qquad 
 = -2 \Omega^{-(d-4)} \tn^a \sigma^{BD}
                        \left(
                              C^f{}_{D a} {\bar S}_{Ef} 
                              - C^u{}_{DE} {\bar S}_{a u} 
                        \right) \sigma^{CE}C^u{}_{BC} + O(\Omega) 
\non \\ && \qquad 
 = -2 \Omega^{-(d-4)} \sigma^{BD}C^C{}_{Du}C^u{}_{BC} 
     + O(\Omega) 
\non \\ && \qquad 
 = O(\Omega) \,, 
\eena 
where the capital letters describe coordinate components with respect 
to $x^A$, and where in the second line we have used 
$\tn^a = (\partial /\partial)^a + O(\Omega^2)$ and 
${\bar S}_{ab} = \sigma_{AB}(\dd x^A)_a(\dd x^B)_b + O(\Omega^2)$, 
and in the last line, $C^C{}_{Du}= O(\Omega^{(d-2)/2})$.  

\medskip 
Next, we consider the second term of the right-hand side of 
eq.~(\ref{eq:toward:A10}),   
\bena 
 && - 8 \Omega^{-(d-3)} \tn^{[a}q^{b]d}  
                          \tC_{aedf} \tn^f q^{ce}C^g{}_{bc} \tell_g 
\non \\
 && \qquad 
 = 4 \Omega^{-(d-3)} q^{ce}q^{bd} \tC_{dfea} \tn^f \tn^a \tnabla_b \tell_c
  + 4 \Omega^{-(d-3)} q^{ce}q^{da} \tC_{aedf} \tn^b \tn^f \tnabla_b \tell_c 
\non \\
 && \qquad 
 = 4 \Omega^{-(d-3)}
     \left[ 
           - \tC^{acdf} \tn_a \tn_f 
           + \tC_{ae}{}^d{}_{f}\tn^a\tn^c \tell^e \tn^f 
     \right] \tnabla_c \tell_d \,, 
\label{eq:toward:A10:ii}
\eena 
where we have used 
${\bar \nabla}_a \tell_b = -\Omega \tell_a \tell_b$ and $q^{ab}\tell_a = 0$, 
and where in the second line we have used that 
\bena 
 q^{ce}q^{bd} \tC_{dfea} \tn^f \tn^a 
 = \tC^{bfca}\tn_f\tn_a - 2\tC^c{}_{aef}\tn^a \tn^b \tell^e \tn^f 
 + \tn^a\tn^b\tn^c \tell^d\tell^e\tn^f \tC_{daef} \,,  
\eena
and 
\bena
  q^{ce}q^{da} \tC_{aedf} \tn^b \tn^f 
  = \tC^c{}_{aef} \tn^a \tn^b \tell^e \tn^f 
  - \tC_{daef} \tn^a \tn^b \tn^c \tell^d \tell^e \tn^f \,, 
\eena 
and the symmetry property of $\tC^a{}_{bcd}$. 
The first term in the third line of the right-hand side of 
eq.~(\ref{eq:toward:A10:ii}) is rewritten as 
\bena 
 && - 4 \Omega^{-(d-3)}\tC^{acdf} \tn_a \tn_f \tnabla_c \tell_d 
\non \\
 && \qquad 
 = - 4 \tnabla_d \left\{
                        \Omega^{-(d-3)}\tC^{acdf} \tn_a \tn_f \tell_c  
                 \right\} 
   + 4 \tell_c \tnabla_d 
                 \left\{
                        \Omega^{-(d-3)}\tC^{acdf} \tn_a \tn_f 
                 \right\} 
\non \\
 && \qquad 
 = - 4 \tnabla_d \left\{
                        \Omega^{-(d-3)}\tC^{acdf} \tn_a \tn_f \tell_c  
                 \right\} 
   + 4 \tell_c \tC^{acdf}\Omega^{-(d-3)}  
                 \left\{
                        (\tnabla_d \tn_a) \tn_f + \tn_a   (\tnabla_d \tn_f) 
                 \right\}  
\non \\
 && \qquad 
 = - 4 \tnabla_d \left\{
                        \Omega^{-(d-3)}\tC^{acdf} \tn_a \tn_f \tell_c  
                 \right\} 
   + 4 \Omega^{-(d-3)}  
       \tC^{acdf}\tell_c\tn_f \tnabla_d \tn_a \,.   
\eena  
Thus, combining all together, we have 
\bena 
 && - 8 \Omega^{-(d-3)} \tn^{[a}q^{b]d}  
                          \tC_{aedf} \tn^f q^{ce}C^g{}_{bc} \tell_g 
\non \\
 && 
  = - 4 \tnabla_d\left(
                        \Omega^{-(d-3)}\tC^{acdf} \tn_a \tn_f \tell_c  
                 \right)  
    + 4 \Omega^{-(d-3)}  
        \left(       
               \tC_{ae}{}^d{}_{f}\tn^a\tn^c \tell^e \tn^f \tnabla_c \tell_d  
                +   
               \tC^{acdf}\tell_c\tn_f \tnabla_d \tn_a 
        \right) \,.   
\eena

\medskip 
Finally we shall show below that the last three terms in the right-hand side 
of eq.~(\ref{eq:50B}) are $O(\Omega)$. 
The first term is 
\ben  
  2\Omega^{-(d-4)} \tC_{abcf}\tn^a\tell^f \tg^{bd} \tg^{ce}{\bar S}_{de} 
  = -4 \Omega^{-(d-4)}\tC^{abcd} n_a \tell_b n_c \tell_d 
    -2 \Omega^{-(d-4)}\tC^{abcd} n_a \tell_d 
       \left(\tg_{bc}- {\bar g}_{bc}\right) \,, 
\een  
and the second terms is 
\ben
 4\Omega^{-(d-3)} \tC_{aedf}\tn^a\tell^e \tn^f\tn^c \tnabla^d \tell_c 
 = 4\Omega^{-(d-4)} \tC_{abcd}\tn_a\tell_b \tn_c \tell_d 
  + O(\Omega) \,, 
\een
where we have used $\tn^e\tnabla_d\tell_e = -\Omega \tell_d + O(\Omega^{d-2})$. 
Combined them, we have 
\bena 
  2\Omega^{-(d-4)} \tC_{abcf}\tn^a\tell^f \tg^{bd} \tg^{ce}{\bar S}_{de} 
+ 4\Omega^{-(d-3)} \tC_{aedf}\tn^a\tell^e \tn^f\tn^c \tnabla^d \tell_c 
 &=& 
 -2\Omega^{-(d-4)} \tC^{abcd} n_a\tell_d \left(\tg_{bc}- {\bar g}_{bc}\right)
\non \\ 
 &=& O(\Omega) \,, 
\eena
where at the last line, we have used $\tg_{bc}- {\bar g}_{bc} 
= O(\Omega^{(d-2)/2})$ and  
\ben
 \tC_{abcd}\tn^d 
 = -\Omega \left(
                  {\bar \nabla}_{[a}\Delta {\bar S}_{b]c} 
                  + C^e{}_{c[a} \tS_{b]e}    
           \right) 
 =O(\Omega^{(d-4)/2}) \,,
\een
which is derived from $ \tC_{abcd}\tn^d = -\Omega \tnabla_{[a} \tS_{b]c}$ 
and $0= {\bar C}_{abcd} {\bar n}^d =-\Omega {\bar \nabla}_{[a} {\bar S}_{b]c}$.
The third term is, 
\bena
 - 8 \Omega^{-(d-4)}\tn^{[a}q^{b]d} C^f{}_{d[a} {\bar S}_{e]f} 
                     q^{ce}C^g{}_{bc} \tell_g  
 &=& -4 \Omega^{-(d-4)}\tn^a 
        {\tilde \gamma}^{BD}{\tilde \gamma}^{CE}C^u{}_{BC} 
                                   C^f{}_{D[a} {\bar S}_{E]f} 
   + O(\Omega) 
\non \\
 &=& -2 \Omega^{-(d-4)}{\tilde \gamma}^{BD} {\tilde \gamma}^{CE}C^u{}_{BC} 
           \tn^a C^F{}_{D a} \sigma_{EF} + O(\Omega) 
\non \\ 
 &=& O(\Omega) \,, 
\eena 
where the capital letters denote coordinate components with respect 
to $x^A$, and where we have used $C^u{}_{ua} = O(\Omega^{(d-2)/2})$ and 
$C^u{}_{\Omega a} = 0$ in the first line, 
$\tn^a {\bar S}_{af} = (\dd \Omega)_f + O(\Omega^{d/2})$, 
$C^\Omega{}_{ab}= O(\Omega^{(d-2)/2})$, and ${\bar S}_{EF} = \sigma_{EF}$ 
in the second line, and used 
$n^aC^F{}_{Da}\sigma_{EF} = O(\Omega^{d/2})$ in the third line.  

%---------------------------------------------------------------------------

\end{document}